%% file: QSO_PSF_energetics.tex
\documentclass{aa}
 \usepackage[T1]{fontenc}
 \usepackage[varg]{txfonts}
 \usepackage{multirow}
 \usepackage{color}
 \usepackage{times}
 \usepackage{amssymb}
 \usepackage{amsmath}
 \usepackage{booktabs}
 \usepackage{upgreek}
 
 \newcommand{\FeII}{\ion{Fe}{ii}}
 \newcommand{\OIII}{\ion{O}{iii}}

 \newcommand{\CIV}{\ion{C}{iv}}
 \newcommand{\MgII}{\ion{Mg}{ii}}
 \newcommand{\Hb}{\mbox{H$\beta$}}

 \newcommand{\QDeb}{\textsc{qdeblend${}^{\mathrm{3D}}$}}

\bibpunct{(}{)}{;}{a}{}{,}

\begin{document}

\title{Large-scale outflows in luminous QSOs revisited}
\subtitle{The impact of beam smearing on AGN feedback efficiencies}
\author{B.~Husemann\inst{\ref{inst1}}\thanks{ESO Fellow, eMail: bhuseman@eso.org} \and J.~Scharw\"achter\inst{\ref{inst2},\ref{inst3}} \and V.~N.~Bennert\inst{\ref{inst4}} \and 
V.~Mainieri\inst{\ref{inst1}} \and 
J.-H.~Woo\inst{\ref{inst5}} \and 
D.~Kakkad\inst{\ref{inst1}}}
\institute{
European Southern Observatory, Karl-Schwarzschild-Str. 2, 85748 Garching b. M\"unchen, Germany\label{inst1}\and 
LERMA, Observatoire de Paris, PSL, CNRS, Sorbonne Universit\'es, UPMC, F-75014, Paris, France\label{inst2} \and 
Gemini Observatory, Northern Operations Center, 670 North A'ohoku Place, Hilo, HI, 96720, USA\label{inst3}\and
Physics Department, California Polytechnic State University, San Luis Obispo, CA 93407, USA\label{inst4} \and 
Department of Physics and Astronomy, Seoul National University, Seoul, 151-742, Korea\label{inst5}}

\authorrunning{Husemann et al.}
\abstract{Feedback from Active Galactic Nuclei (AGN) is thought to play an important role in quenching star formation in galaxies. However, the efficiency with which AGN dissipate 
their radiative energy into the ambient medium remains strongly debated.}
{Enormous observational efforts have been made to constrain the energetics of AGN feedback by mapping the kinematics of the ionized gas on kpc scale. We 
study how the observed kinematics and inferred energetics are affected by beam smearing of a bright unresolved narrow-line region (NLR) due to seeing.}
{We analyse optical integral-field spectroscopy of a sample of twelve luminous unobscured QSOs ($0.4<z<0.7$) initially presented by \citet{Liu:2014}. The 
point-spread function (PSF) for the observations is directly obtained from the light distribution of the broad \Hb\ line component. Therefore, we are able to compare the 
ionized gas kinematics and derived energetics of the total, truly spatially extended, and unresolved [\OIII] emission. }
{We find that the spatially resolved [\OIII] line width on kpc scales is significantly narrower than the one before PSF deblending. The extended NLRs (ENLRs) appear intrinsically offset 
from the QSO position or more elongated which can be interpreted in favour of a conical outflow on large scales while a spherical geometry cannot be excluded for the unresolved NLR. We find 
that the kinetic power at 5\,kpc distance from the spherical model by \citet{Liu:2013b} is reduced by two orders of magnitude for a conical outflow and one 
order of magnitude for the unresolved NLR after PSF deblending. This reduced kinetic power corresponds to only 0.01--0.1\,per cent of the 
bolometric AGN luminosity. This is smaller than the 5-10\,per cent feedback efficiency required by some cosmological simulations to reproduce the massive galaxy population. The injected 
momentum fluxes are close or below the simple radiation-pressure limit $L_\mathrm{bol}/c$ for the conical outflow model for the NLR and ENLR when beam smearing is considered.}
{Integral-field spectroscopy is a powerful tool to investigate the energetics of AGN outflows, but the impact of beam smearing has to be taken into account in the high contrast 
regime of QSOs. For the majority of observations in the literature, this has not been addressed carefully so that the incidence and energetics of presumed kpc-scale AGN-driven outflows 
still remain an unsolved issue, from an observational perspective.}
\keywords{ISM: jets and outflows - Galaxies: active - quasars: emission lines - Techniques: imaging spectroscopy}
\maketitle

\section{Introduction}
Feedback from Active Galactic Nuclei (AGN) has become a key ingredient in numerical simulations and semi-analytic models of galaxy evolution to suppress star formation at the highest 
stellar masses, which appears necessary to recover the properties of the local galaxy population \citep[e.g.][]{Matteo:2005,Bower:2006,Croton:2006,Somerville:2008,Schaye:2015}. However, the 
mechanism(s) with which the released energy of AGN is dissipated to the surrounding interstellar medium of the host galaxy is poorly constrained by observations so far. One popular scenario 
for AGN feedback is a large-scale outflow where the AGN energy is sufficient to expel a large fraction of the gas from the host galaxy \citep[e.g.][]{Silk:1998}. Thereby, the AGN is reducing the 
available gas reservoir and the star formation activity becomes greatly suppressed. 

The existence of high-velocity AGN-driven gas outflows has been confirmed by X-ray observations of ultra-fast outflowing material in the circumnuclear region 
\citep[e.g.][]{Tombesi:2010,Gofford:2015}. Also, broad-absorption line (BAL) AGN display outflowing gas with velocities of a few 1000\,$\mathrm{km}\,\mathrm{s}^{-1}$ in UV absorption lines like 
[\CIV] and \MgII\ \citep[e.g.][]{Reichard:2003a,Trump:2006,Gibson:2009}. The extent of the outflows is usually not directly constrained from those observation due to a lack of spatial resolution. A 
remedy to this dilemma is provided through the (extended) narrow-line region (E)NLR, which corresponds to the gas ionized by the AGN on tens of pc to tens of kpc scales 
\citep[e.g.][]{Pogge:1988b,Capetti:1996,Bennert:2002,Schmitt:2003a,Hainline:2014,Keel:2015}. The demarcation between NLR and ENLR is arbitrary given that the ionization mechanism is the 
same, but a transition radius at $\sim$1\,kpc has been used \citep[e.g.][]{Unger:1987}. The bright [\OIII]\,$\lambda\lambda4960,5007$ doublet line ([\OIII] hereafter) is mainly used at optical 
wavelengths to trace the kinematics in the NLR close to the AGN. It is well-known that this line tends to be systematically asymmetric with a blue wing that is interpreted as a genuine signature for 
an extended outflow \citep[e.g.][]{Heckman:1981,Boroson:2005,Komossa:2008,Mullaney:2013}. These high-velocity outflows are well resolved in very nearby Seyfert galaxies on $<1$\,kpc scales 
\citep[e.g.][]{Crenshaw:2000,Rice:2006,Storchi-Bergmann:2010,Fischer:2013}. 

Currently, large efforts are being made to investigate the properties of large-scale high-velocity outflows in the most luminous AGN, i.e. quasi-stellar objects (QSOs). They are expected to 
show the strongest outflows if these are driven by the AGN radiation \citep[e.g.][]{Silk:1998,King:2003,Hopkins:2010,Faucher-Giguere:2012}.  With the advent of optical and near-IR long-slit or 
integral-field unit (IFU) spectrographs on 8m class telescopes, the ENLR kinematics has been mapped on kpc scale for luminous QSOs at low redshift $z<1$ 
\citep[e.g.][]{Fu:2009,Greene:2011,Rupke:2011,Villar-Martin:2011b,Greene:2012,Husemann:2013a,Liu:2013b,Harrison:2014,Liu:2014,McElroy:2015,Liu:2015,Humphrey:2015,Villar-Martin:2016,Karouzos:2016} and 
high redshift $z>1$ \citep[e.g.][]{CanoDiaz:2012,Harrison:2012,Brusa:2015,Cresci:2015,Carniani:2015}. The majority of those studies report outflows on several kpc scales in almost all QSOs as 
indicated by broad and/or blue-shifted [\OIII] emission lines.

Many of the spectroscopic QSO observations are focussed on obscured (type II) QSOs. Obscured QSOs lack the bright point-like power-law continuum of the accretion disc and the emission of the 
broad-line region (BLR) that are prominent in unobscured (type I) QSOs. This difference has been explained by the inclination of a toroidal-like obscuring structure with respect to our line-of-sight 
in the unification model of AGN \citep[e.g.][]{Antonucci:1993}. In this model the NLR is located outside the obscuring structure and can be seen in both types of QSOs. Given the high gas density and 
radiation field close to the AGN, the  [\OIII] emission lines from the NLR on scales of $\ll$1\,kpc can outshine the ENLR on host galaxy scales by a factor of a few depending on the size of 
the ENLR and physical resolution of the observation. Therefore, we define the NLR and ENLR as the spatially unresolved and resolved emission, respectively, for the purpose of our study.

It is therefore crucial to characterize the point-spread function (PSF) for QSO observations in order to separate the emission from the compact NLR and the contributions from the ENLR. The ability to 
achieve such a separation depends strongly on the spatial resolution. Characterizing the PSF is particularly challenging for spectroscopic observations of obscured QSOs, because the slit or IFU does 
usually do not cover a star simultaneously. In these cases, an approximation of the PSF and its shape may be obtained from 
acquisition images 
\citep[e.g.][]{Hainline:2013,Hainline:2014,Humphrey:2015} or standard star observations \citep[e.g.][]{Liu:2013,Liu:2014} taken close in time to the science observations.  However, the actual PSF for 
the science observation can still be different due to time variability of the seeing and the tracking error of the telescope for the significantly 
longer science exposures. 

IFU spectroscopy of unobscured QSOs provides a way to reconstruct the PSF directly from the science data assuming that broad Balmer lines from the BLR are intrinsically unresolved
\citep[e.g.][]{Jahnke:2004}. This technique was applied to various IFU observations of unobscured QSOs 
\citep{Sanchez:2004a,Christensen:2006,Husemann:2008,Husemann:2013a,Husemann:2014,Carniani:2015,Herenz:2015,Liu:2015} enabling the study of line diagnostics and kinematics across the host 
galaxy without the apparent contamination of the bright unresolved NLR. Based on a large sample of luminous unobscured QSOs at $z<0.3$ and luminous obscured QSOs, 
\citet{Husemann:2013a}, \citet{Villar-Martin:2016} and \citet{Karouzos:2016} reported a lack of high-velocity outflows on kpc scales after deblending the unresolved NLR and ENLR. 
This appears to be in direct contradiction to the result of various other groups for luminous obscured QSOs \citep[e.g.][]{Greene:2011,Liu:2013,Liu:2014,Harrison:2014,McElroy:2015}. However, those 
studies did not separate kinematics of the NLR and ENLR given the difficulty to constrain the PSF. It is therefore unclear whether beam smearing, differences in the QSO feedback efficiency or even 
differences in the unobscured and obscured QSOs sample selection are causing these discrepant conclusions.

In this paper, we systematically investigate the effect of the beam smearing on the measured kpc-scale kinematics of the [\OIII] lines. In particular, \citet{Liu:2013b} reported very large 
mass outflow rates and kinetic power from IFU spectroscopy of the [\OIII] line for a sample of luminous obscured QSOs at redshift $0.4<z<0.7$ . While the PSF for these observation cannot be 
reconstructed, the authors presented also a matched sample of unobscured QSO in \citet{Liu:2014} for which the BLR can be used as a PSF tracer. Here, we re-reduce and re-analyse the dataset of 
luminous unobscured QSOs from \citet{Liu:2014} and compare the results with and without deblending the contribution from the unresolved NLR and ENLR. Thereby, we can verify how much the results on 
the ENLR geometry, the large-scaled ionized gas kinematics and associated AGN feedback efficiency are affected by beam smearing.

The paper is organized as follows. In Section~\ref{sect:sample} we describe the IFU data reduction and analysis including  our QSO-host galaxy deblending scheme and emission-line measurements. 
This is followed by a comparison of various parameters on the extended ionized gas measurements before and after deblending the point-like and extended emission (Sect.~\ref{sect:comparison}). From 
the measured quantities we compute outflow energetics for two different outflow models in Sect.~\ref{sect:feedback}. We then discuss our results with previous observations and expectations for AGN 
feedback scenarios (Sect.~\ref{sect:discussion}). Finally, we close with a summary and our main conclusions in Sect.~\ref{sect:conclusion}. Throughout the paper we assume a concordance cosmological 
model with $H_0=70\,\mathrm{km}\,\mathrm{s}^{-1}\,\mathrm{Mpc}^{-1}$, $\Omega_{\mathrm{m}}=0.3$, and $\Omega_\Lambda=0.7$.

\begin{figure*}
\centering
\includegraphics[width=\textwidth]{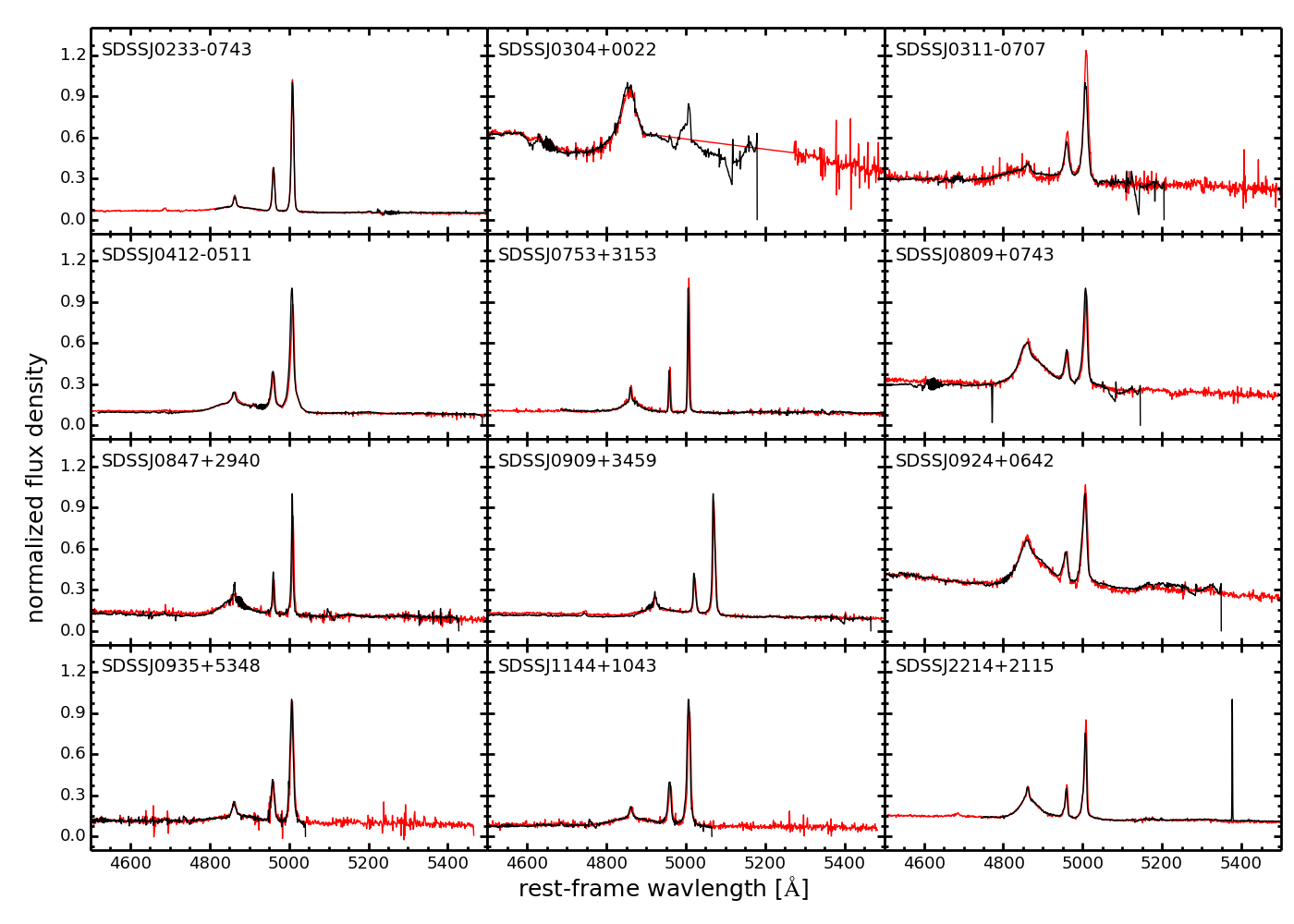}
\caption{Comparison of the GMOS $3''$ aperture spectra (black line) with the corresponding SDSS spectrum (red line). The wavelength range is limited to the rest-frame wavelength range between 4500 
and 5500$\AA$. An absolute photometric calibration of the GMOS data based on the median of the ratio of spectra over the common wavelength range. The spectra are normalized in flux density 
so that the peak in the [\OIII] $\lambda5007$ line is set to one in the SDSS spectrum. A large part of the SDSS spectrum of SDSSJ0304+0022 is masked as bad which is seen as the linear interpolated 
region. In general, the relative GMOS flux calibration across the wavelength range is consistent with the SDSS spectra at a $<$10 per cent level.}
\label{fig:GMOS_spec}
\end{figure*}

\section{Integral-field spectroscopy of luminous QSOs}\label{sect:sample}
\subsection{The QSO sample and IFU data reduction}
The QSO sample and optical integral-field observation we focus on in this paper are presented by \citet{Liu:2014}; we briefly recap the main characteristics of the sample. 
The 12 QSOs were selected from the \citet{Shen:2011} catalogue to have (i) a minimum [\OIII] luminosity of $L_\mathrm{[\OIII]}>10^{42.7}\,\mathrm{erg}\,\mathrm{s}^{-1}$ to be comparable with the 
unobscured QSO selection in \citet{Liu:2013}, (ii) a redshift range of $0.4<z<0.7$, (iii) a 1.4\,GHz radio flux not exceeding $f_\mathrm{1.4GHz}<10\,\mathrm{mJy}$ in the NVSS or FIRST radio surveys 
to exclude radio-loud QSOs, and (iv) a high [\OIII] equivalent width. In Table~\ref{tab:sample}, we list some characteristic parameters of the sample mainly taken from \citet{Liu:2014}, but we also 
compute additional parameters from the data itself, i.e. the broad \Hb\ line luminosity ($L_{\mathrm{H}\beta}$) and the continuum luminosity at 5100\AA\ ($L_{5100}$).

Observations of the QSO sample were taken with the Gemini Multi-Object Spectrograph \citep[GMOS:][]{Allington-Smith:2002} in IFU-mode at the Gemini-North telescope as part of programme GN-2012B-Q-29 
(PI: G. Liu). We retrieved the raw data and corresponding calibrations from the GEMINI science archive after the data became publicly available. The two-slit mode of the GMOS IFU provides a 
$5\arcsec\times7\arcsec$ target field-of-view (FoV) that is contiguously sampled with 1000 hexagonal lenslets of $0\farcs2$ in diameter. Additionally, 500 lenslets are packed into a $5''\times3.5''$ 
FoV about $1\arcmin$ offset from the primary IFU field to simultaneously monitor the sky. The spectral range was chosen such that \Hb\ and [\OIII] lines are simultaneously covered. Two different 
setups with the R400-G5305 grism ($R\sim2000$) in the $i$ band are necessary to capture those important lines considering the redshift range in the sample and to avoid that the lines fall in one of  
the gaps between the three CCDs. Two 1620\,s exposure were obtained for each QSO in the sample.

For the data reduction, we use the IFU data reduction package developed and extensively tested for the Calar Alto Large Integral Field Area (CALIFA) survey 
\citep{Sanchez:2012a,Husemann:2013b,Garcia-Benito:2015}. Since CALIFA uses the fibre-based IFU spectrograph PMAS \citep{Roth:2005}, all reduction steps are almost identical except 
that the GMOS IFU samples the FoV contiguously and that the data is spread over three independent CCDs. The data reduction work-flow consists of standard tasks such as bias subtraction, cosmic-ray 
masking with PyCosmic \citep{Husemann:2012a}, fibre tracing and fibre profile fitting using the continuum lamp exposure, flexure correction, optimal fibre extraction, wavelength calibration using the 
attached arc lamp exposure, and relative wavelength-dependent fibre transmission correction using a continuum lamp. One important difference with respect to the data reduction of \citet{Liu:2014} is 
that we re-sample the data into a datacube with $0\farcs2$ rectangular spaxels. Over-sampling the native data resolution with just two exposures does not provide additional information and would 
degrade the S/N per final spaxel. For the re-sampling, we assume that the hexagons effectively collect light within a circular aperture of $0.2''$ diameter and apply the ''drizzle'' resampling scheme 
\citep{Fruchter:2002} to construct the final datacubes. During this re-sampling step, we simultaneously correct for the effect of atmospheric dispersion by shifting the sample grid to account for the 
continuous shift in the relative position along wavelength. 

Standard star observations are reduced in the same way as the science data to perform a relative spectrophotometric flux calibration along wavelength. Following \citet{Liu:2014}, we retrieve 
Sloan Digital Sky Survey \citep[SDSS,][]{York:2000} spectra from DR10 \citep{Ahn:2014} and compare the GMOS spectra with the SDSS ones to anchor our absolute flux calibration. Here, we simply extract 
spectra within a $3''$ diameter centred on the QSO and compare it directly with the SDSS DR10 spectra since they were already re-scaled in flux to account for the  aperture fibre losses. Then we 
determine the photometric scale factor compared to the SDSS spectra as the median of the ratio between the two spectra. We show the SDSS and the matched aperture GMOS spectra in 
Fig.~\ref{fig:GMOS_spec}. Although the spectrophotometric calibration of SDSS spectra are considered very accurate, the GMOS and SDSS data are taken a few years apart so that intrinsic 
variability of AGN in the continuum and broad lines will lead to systematic uncertainties in our adopted absolute photometric calibration.

\begin{table*}
\centering
\caption{Basic sample characteristics.}
\label{tab:sample}
 \input{table1.tex}
\end{table*}

\begin{figure}
\centering
\resizebox{\hsize}{!}{\includegraphics{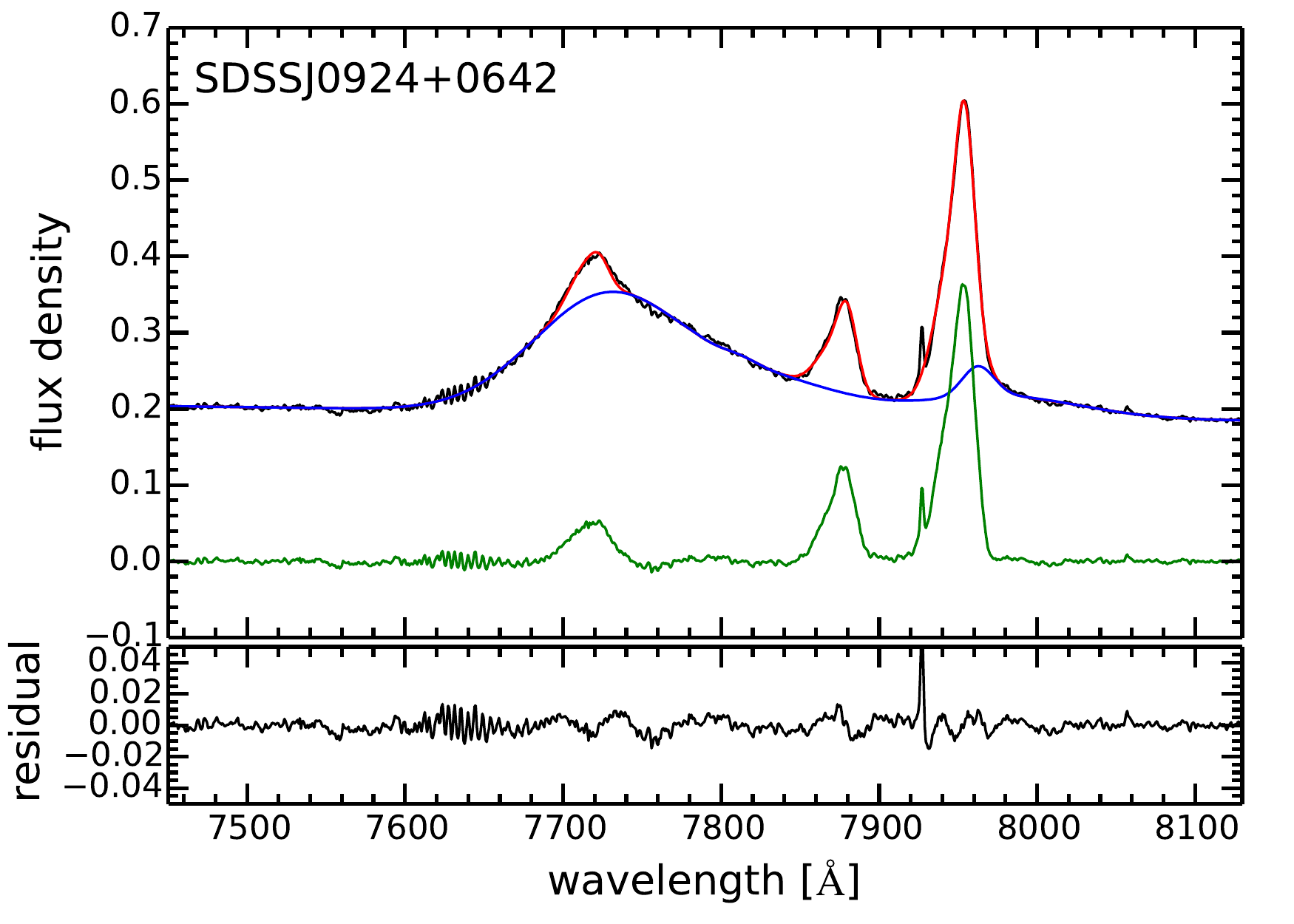}}
\caption{Example of the broad \Hb\ and \FeII\ emission-line subtraction without QSO-host deblending for SDSS~J0924$+$0642. The observed spectrum (black line) and our best-fit model (red line) are 
shown in the upper panel. The best-fit model consisting of the broad \Hb\ and \FeII\,$\lambda\lambda4948,5017$ plus continuum is represented by the blue line with the corresponding residual spectrum 
of the narrow \Hb\ and [\OIII]\,$\lambda\lambda4960,5007$  indicated by the green line. The residuals of the total model are shown in the panel below. Details of the assumed model are given in the 
main text.}
\label{fig:EM_fit}
\end{figure}

\begin{figure}
\centering
\resizebox{\hsize}{!}{\includegraphics{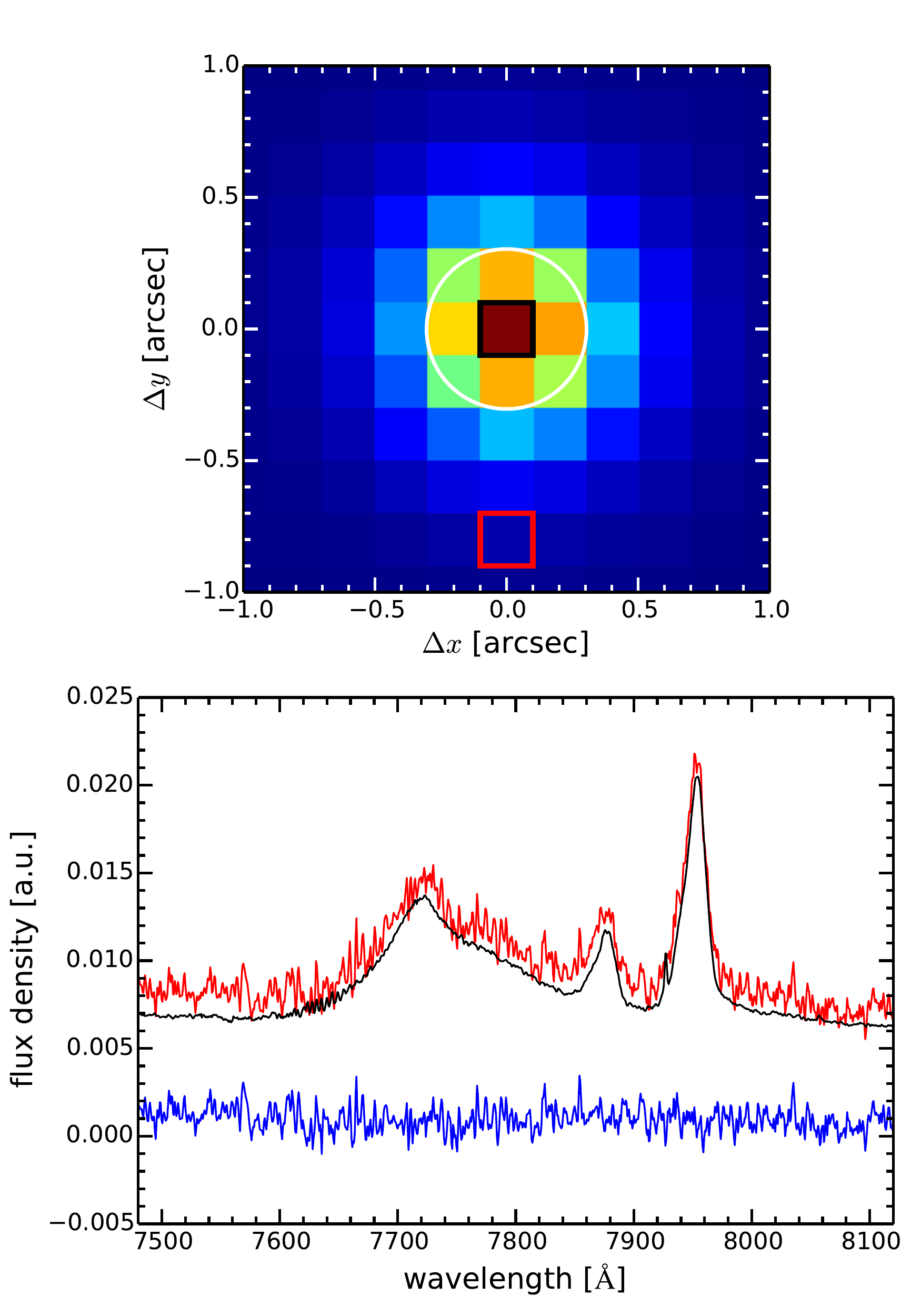}}
\caption{{\it Upper panel:} Example point-spread function (PSF) estimated from the intensity of the broad \Hb\ line for SDSS~J0924$+$0642. The white circle indicates the FWHM of the seeing. We 
highlight two spaxels with a black and red square for which we show the spectra from the datacube in the lower panels. The red spaxel is $0\farcs8$ (5.8\,kpc) away from the QSO position.
{\it Lower panel:} Spectra from the two spaxels highlighted in the upper panel. The central spaxel spectrum is scaled in to match in the integrated broad \Hb\ flux of outer spectrum. The difference 
between the spectra are indicated by the blue line and shows that both spectra are identical in shape except of a constant continuum offset across the wavelength range. Any apparent emission line 
contribution in the red spaxel is simply due to beam smearing of an unresolved source even for the forbidden [\OIII] line from the NLR.}
\label{fig:deblend}
\end{figure}

\subsection{Spatially-resolved [\OIII] emission-line analysis}
The first analysis step is usually to create emission-line maps and parameters from the individual spaxel of the datacube. A generic feature of ground-based observations 
is that the signal from the source is spatially smeared due to the seeing, so that the spaxels may not be independent in their information content. In particular, luminous unobscured QSOs are subject 
to this effect since the light from the accretion disc and BLR are emitted from a very compact region but which can be as bright as the entire host galaxy. Nevertheless, the beam smearing affects the 
bright narrow lines like the [\OIII]$\lambda\lambda4960,5007$ lines in unobscured and obscured QSOs in exactly the same way. The reason for this is that the classical compact NLR on $\ll 
1\,\mathrm{kpc}$ 
scales seen for both types of AGN becomes quickly unresolved with increasing redshift and can outshine any ionized gas emission of the ENLR, which extends over several kpc for luminous QSOs 
\citep[e.g.][]{Bennert:2002,Husemann:2014,Hainline:2014,Keel:2015}, depending on the size and contrast ratio.  

Here, we specifically want to test how much the light from the unresolved NLR blends with the ENLR in IFU observations, altering the spatially resolved line profiles and biasing the derived 
quantities. Unobscured QSOs are ideal for this purpose since the broad emission-lines from the unresolved BLR provide an intrinsic measurement for the point-spread function (PSF) of a given 
observation \citep[e.g.,][]{Jahnke:2004}. This allows to accurately deblend spatially unresolved, NLR or QSO, and resolved, ENLR or host galaxy emission in an empirical way. We will refer to this 
process as a NLR-ENLR deblending or QSO-host galaxy deblending.  To make a fair comparison, we characterize the [\OIII] $\lambda\lambda4960,5007$ doublet line profile spaxel by spaxel in a consistent 
way before and after applying NLR-ENLR deblending as described below.

\begin{figure}[h!]
\centering
\includegraphics[width=0.42\textwidth]{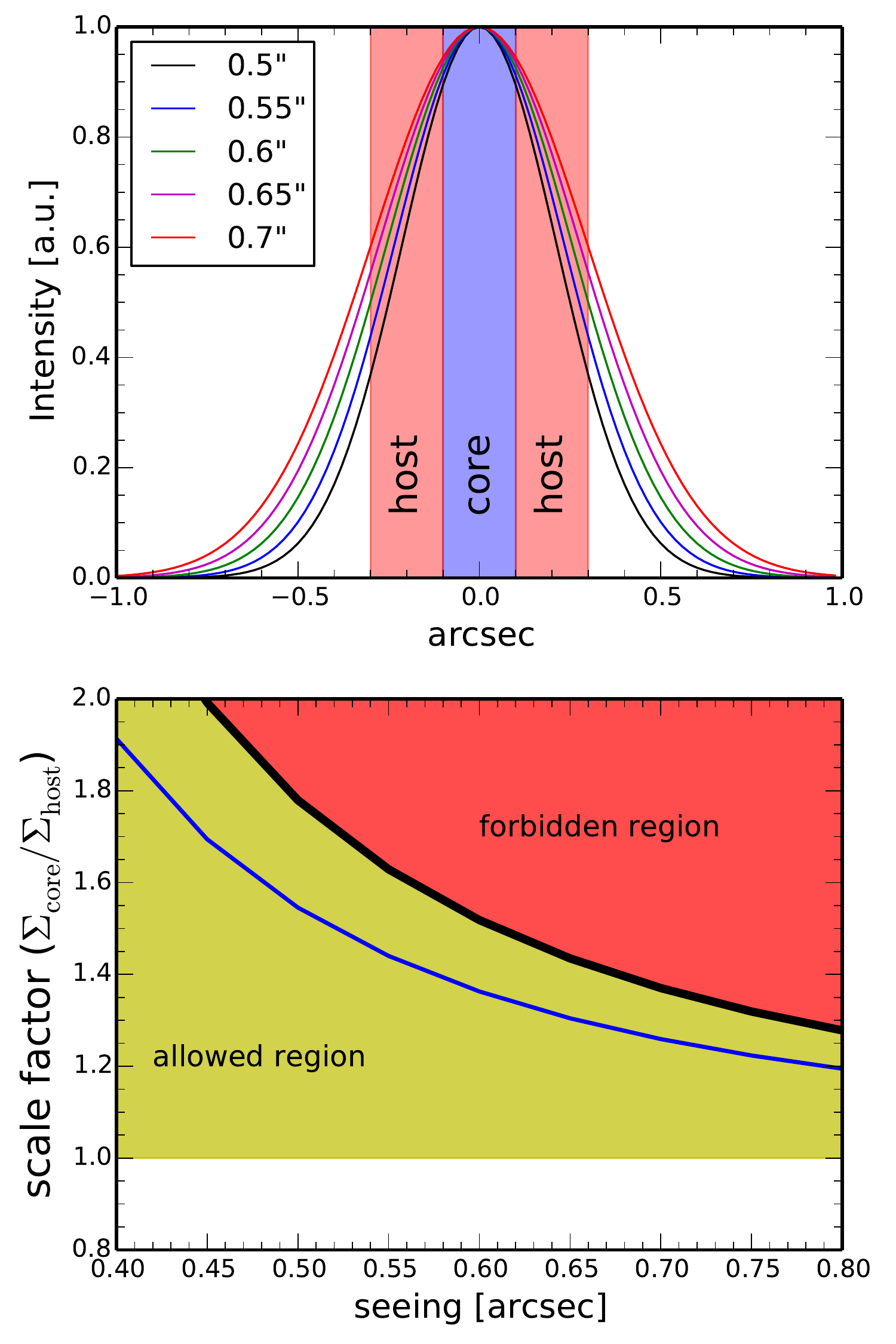}
\caption{{\it Upper panel:} Illustration of the intensity profile for various seeing conditions assuming a Gaussian shape. The region of the central spaxel (core) and the adjacent host 
galaxy spaxel (host) are indicated by shaded areas for the GMOS sampling.
{\it Lower panel:} The corresponding scale factors for a point-source as a function seeing is shown by the black line. Higher scale factors would imply surface brightness 
distributions steeper than point-like sources (red shaded area). The assumed scale factor we use in our case for the QSO-host deblending is chosen to be just below the allowed limit for a 
conservative 
estimate of the extended flux.}
\label{fig:PSF_scaling}
\end{figure}

\subsubsection{Mapping the total [\OIII] line profile}
To characterize the spatially resolved [\OIII] emission-line profile, we follow the algorithm of \citet{Liu:2013b} which consists of three basic steps: 1) removal of \FeII\ and broad \Hb\ 
emission, 2) multi-component modelling of the [\OIII] doublet line, and 3) non-parametric line shape measurements based on the best-fit model.  The first step is achieved by modelling the 
QSO spectrum with a set of Gaussian profiles to separate the various emission line components in the spectral range (see Fig.~\ref{fig:EM_fit}). Then, we create a best-fit model only for the 
broad \Hb\ and \FeII\ line components as well as the local AGN power-law continuum. This template spectrum is subtracted from each spaxel after proper matching in flux. The residual is a pure 
narrow \Hb\ plus [\OIII] emission-line datacube. According to \citet{Liu:2013b}, we model the [\OIII] doublet lines as a superposition of up to three independent Gaussian systems coupled in their 
intrinsic flux ratio \citep[1:3,][]{Storey:2000}, redshift and line dispersion.  A Levenberg-Marquardt minimization algorithm with reasonable starting values is applied to 
determine the best-fit parameters per spaxel. Any model with an increased number of free parameters will provide a better fit. Based on a statistical $F$-test we decide whether the increased number 
of free parameters significantly improved the $\chi^2$ in excess of what is expected from statistical fluctuations.  From the [\OIII] line shape of the best-fit model we directly compute 
non-parametric parameters such as the integrated flux $f_\mathrm{[\OIII]}$, the median line velocity $v_\mathrm{med}$, the line width at the 80\,per cent quantile of the line flux ($W_{80}$), the 
line asymmetry ($A$) and line kurtosis ($K$)  following the formula presented in \citet{Liu:2013b}. 

\subsubsection{Mapping the ENLR [\OIII] line profile}
We repeat the entire analysis process again, but now replacing step 1) with a spectral QSO-host galaxy deblending scheme to separate the apparently unresolved (NLR) and resolved (ENLR) [\OIII] line 
emission.  The explicit modelling and subtraction of the broad \Hb\ and \FeII\ emission 
lines is not necessary, because those emission lines originate from the BLR  and are intrinsically unresolved emission associated with the QSO spectrum and automatically subtracted during QSO-host 
galaxy deblending process as shown in Fig.~\ref{fig:deblend}. For the QSO-host deblending we adopt an iterative algorithm implemented in the public software package \QDeb\ 
\citep{Husemann:2013a,Husemann:2014}. 

\QDeb\ first re-constructs the PSF of the observations from the strength of the broad emission line, the \Hb\ line in this case. In the first iteration, the brightest spaxel which is dominated by the 
QSO light is scaled according to the PSF and subtracted from each spaxel. The central spaxel contains not only spatially unresolved emission from the QSO, but also a fraction of host galaxy emission 
including the ENLR. The algorithm iteratively removes the host galaxy contribution based on the average surface brightness of the residual host galaxy emission ($\Sigma_\mathrm{host}$ in 
the spaxels around the central QSO spaxel ($\Sigma_\mathrm{core}$) after each iteration. A scale factor $\Sigma_\mathrm{core}/\Sigma_\mathrm{host}$ is applied to scale the brightness towards 
the centre, which clearly depends on the intrinsic surface brightness profile of the extended host galaxy/ENLR emission. It can reasonably vary only between a factor of 1 (constant surface brightness) 
and a factor corresponding to purely unresolved emission depending on the PSF and GMOS 
sampling (see Fig.~\ref{fig:PSF_scaling}).

We choose a factor very close to the scale factor in the limit of a point-like emission, because we have no ancillary information on the exact surface 
brightness distribution of the extended emission on small scales. For all sources we assume a scale factor of about 70\% between the point-like and constant surface brightness value.
This is a conservative choice that avoids significant over-subtraction of extended emission. It also ensures that the process actually converges because the scale factors are below the 
point-like limit in all case, otherwise the QSO spectrum would be oversubtracted. The process usually converges after a few iterations and we choose five iterations for all objects. 
We then repeat the [\OIII] line modelling and the non-parametric measurements in the QSO-subtracted data. The uncertainties of the line profile measurements increase after 
the QSO-host deblending process due to additional uncertainties in the PSF reconstruction and the noise of the subtracted QSO spectrum. In Fig.~\ref{fig:comparison} we present the resulting [\OIII] 
line parameter maps for all the QSOs from the total line profile and from the ENLR only after applying the QSO-host deblending.

\begin{figure*}
\centering
\includegraphics[width=0.99\textwidth]{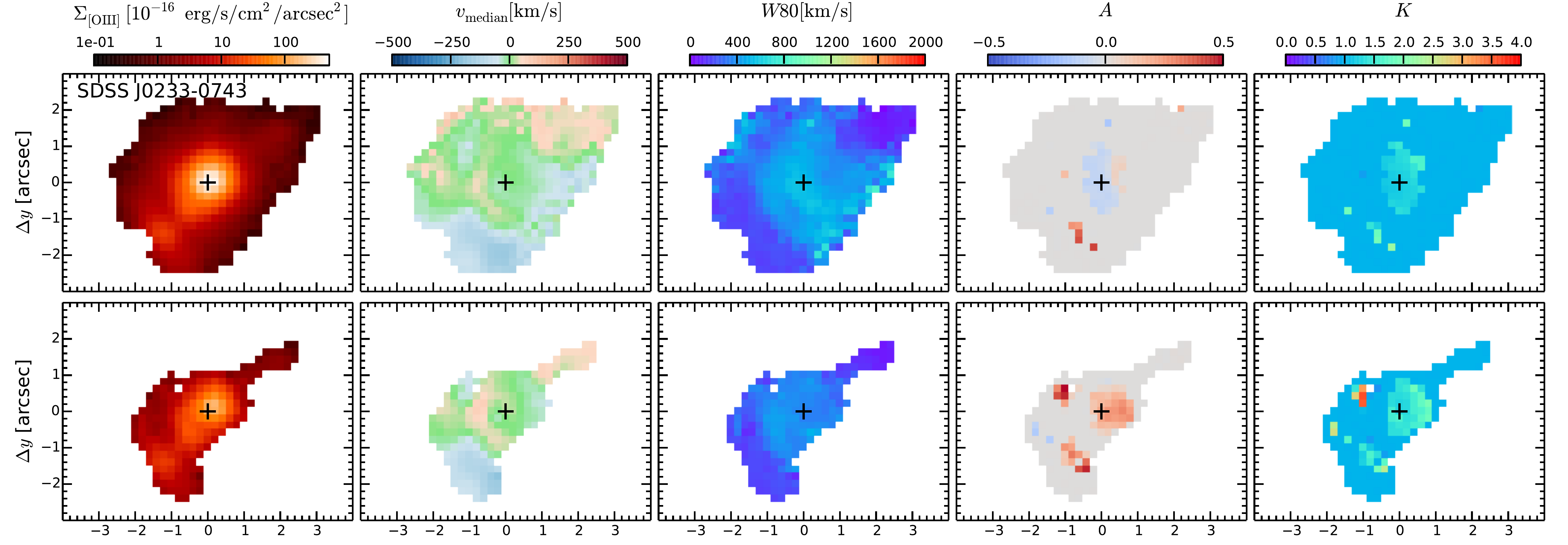}
\includegraphics[width=0.99\textwidth]{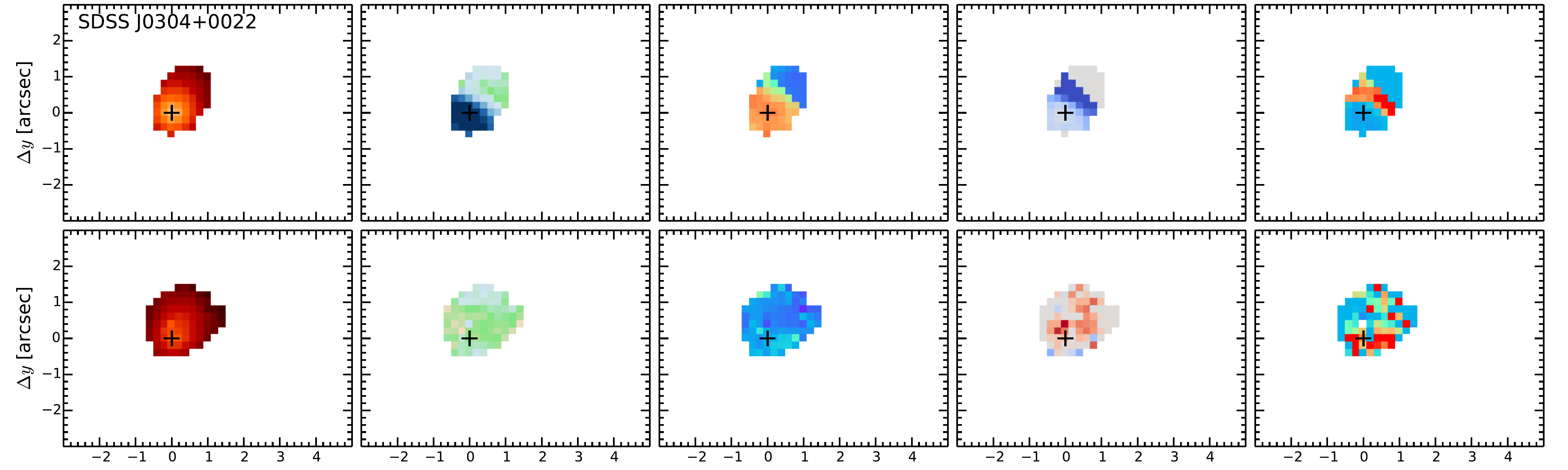}
\includegraphics[width=0.99\textwidth]{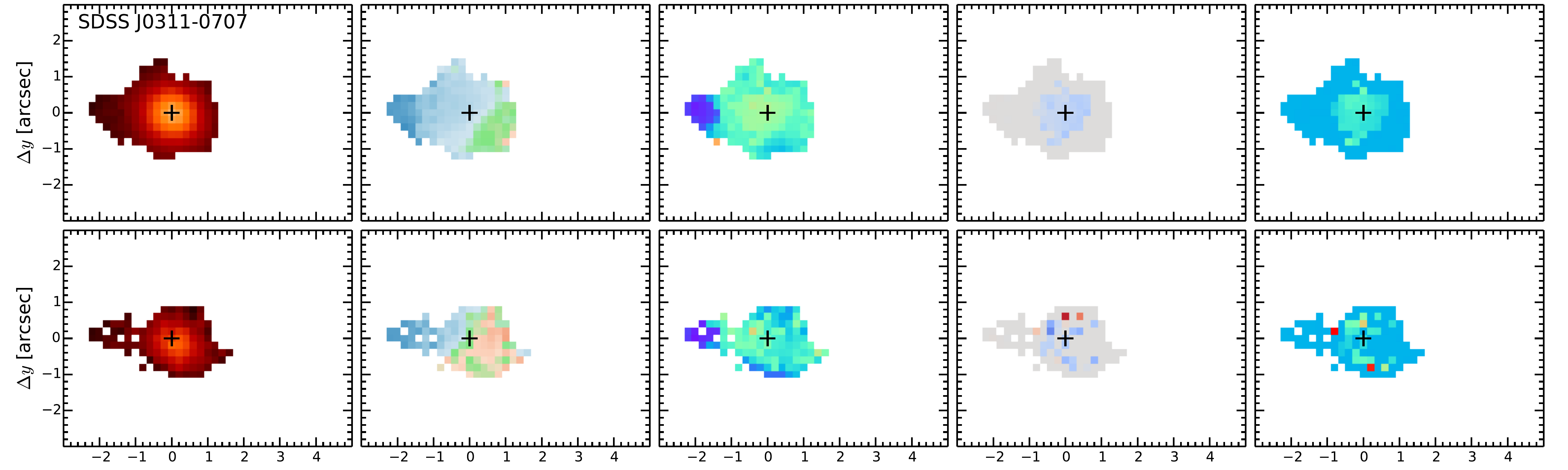}
\includegraphics[width=0.99\textwidth]{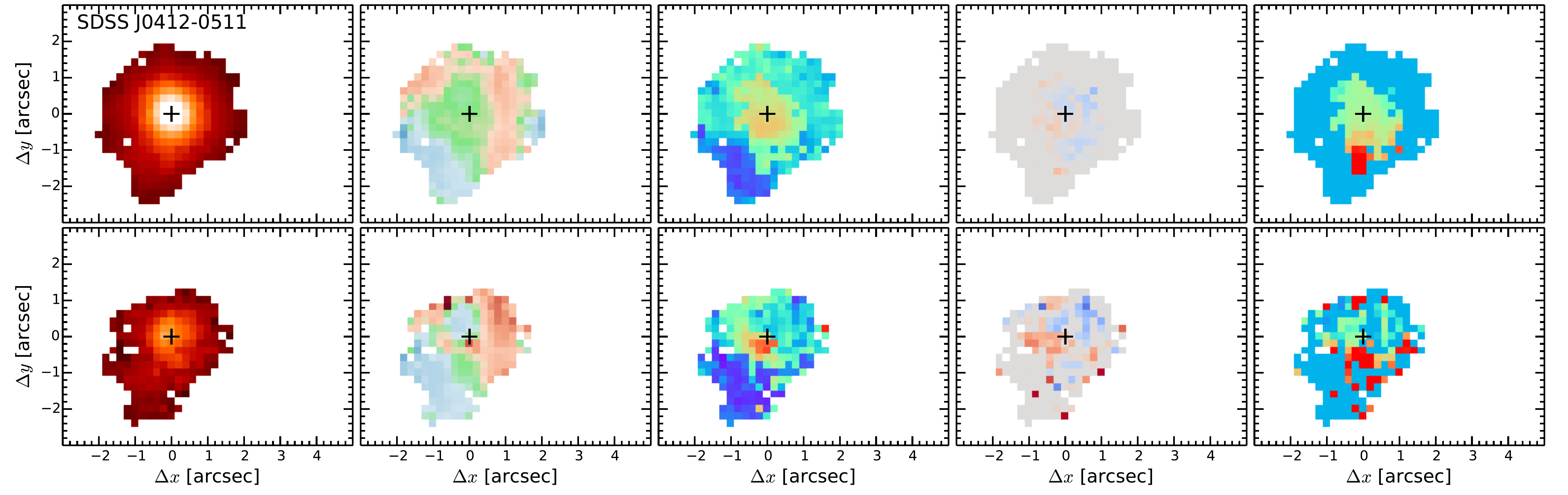}
\caption{Comparison of spatially resolved emission-line parameters across the GMOS FoV before (upper panels)) and after (lower panels) applying the QSO-host galaxy deblending for each QSO. From left 
to right we present the [\OIII] surface brightness distribution ($\Sigma_\mathrm{[\OIII]}$), the median line velocity ($v_\mathrm{median}$), the line width covering 80\,per cent of the line flux 
($W_{80}$), the line asymmetry parameter ($A$) and the kurtosis parameter ($K$).}
\label{fig:comparison}
\end{figure*}
\addtocounter{figure}{-1}

\begin{figure*}
\centering
\includegraphics[width=0.99\textwidth]{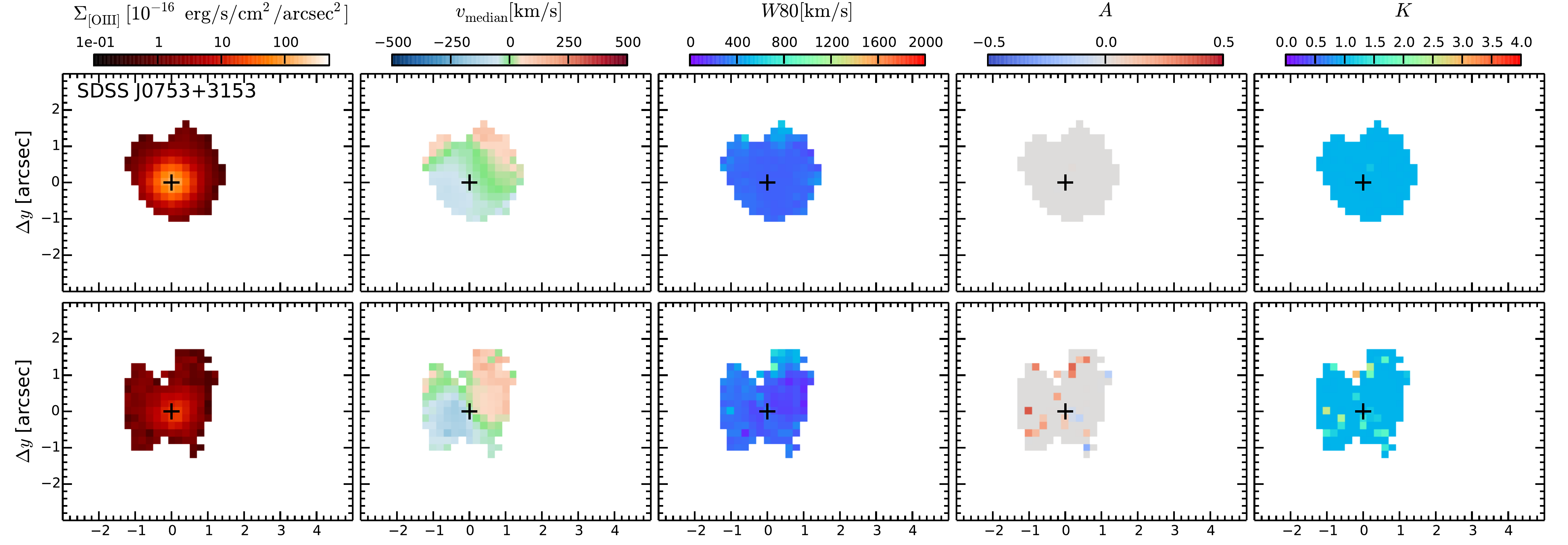}
\includegraphics[width=0.99\textwidth]{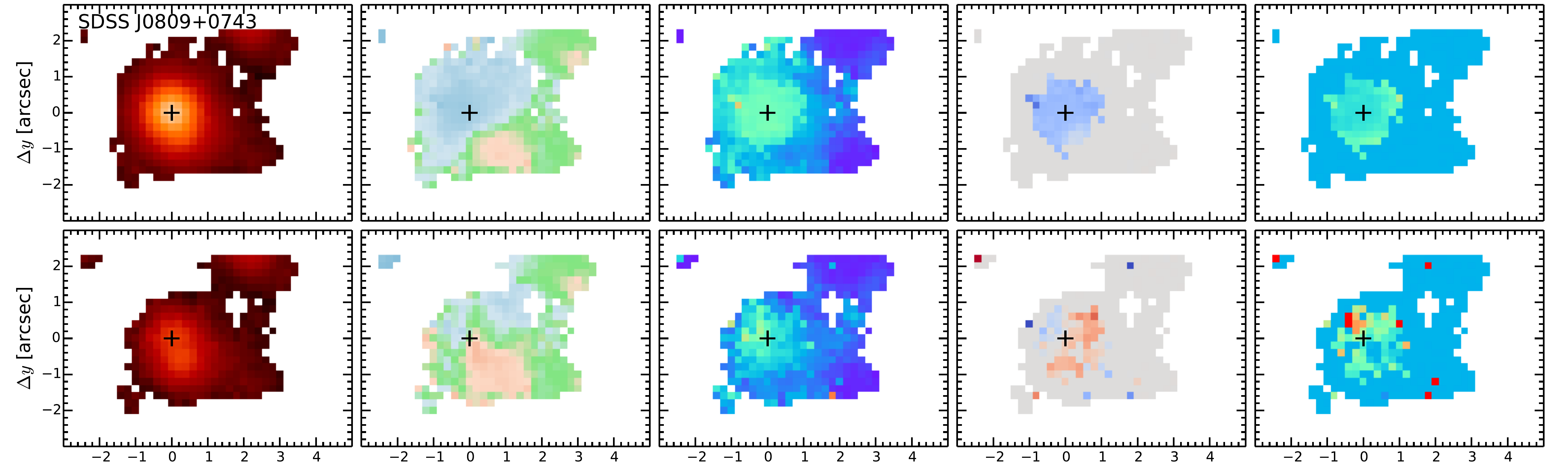}
\includegraphics[width=0.99\textwidth]{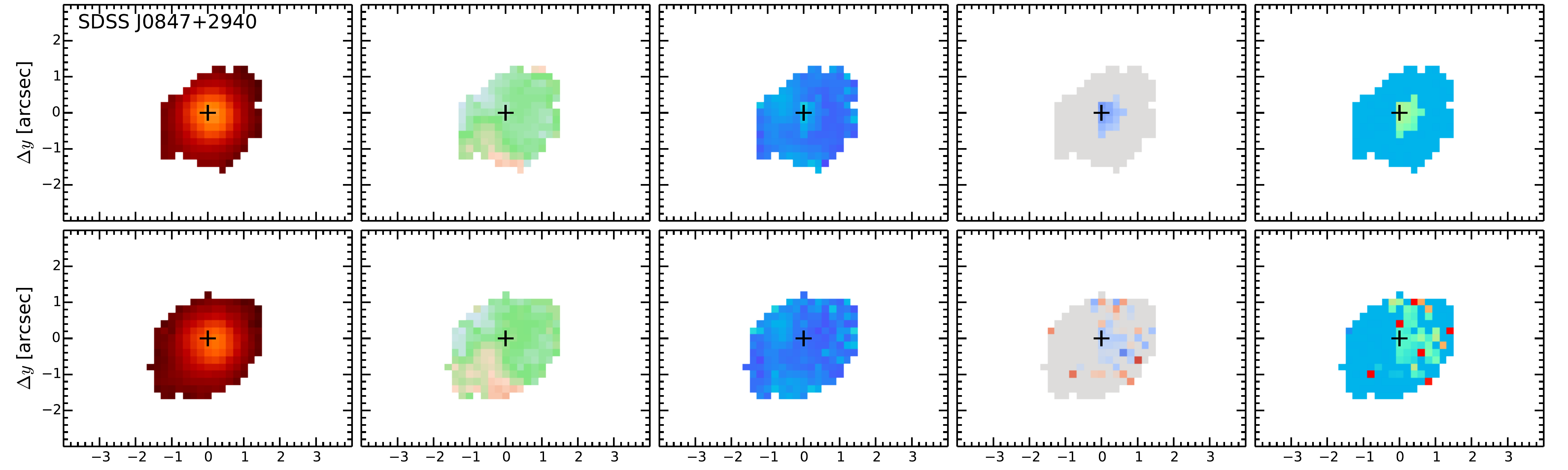}
\includegraphics[width=0.99\textwidth]{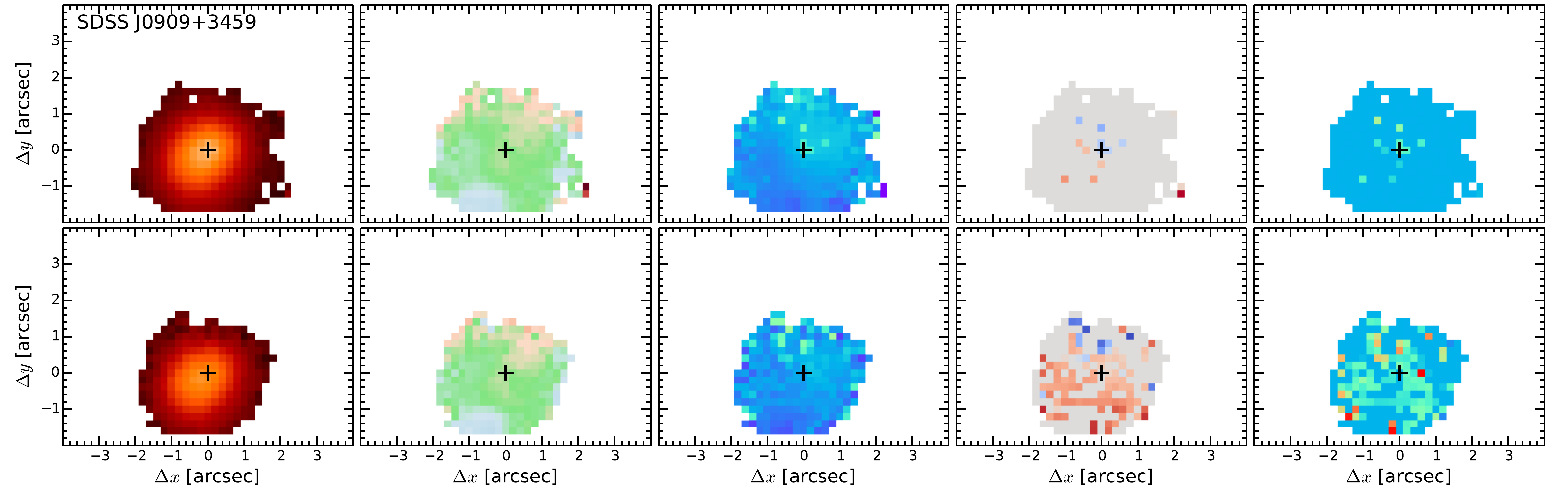}

\caption{continued.}
\end{figure*}
\addtocounter{figure}{-1}

\begin{figure*}
\centering
\includegraphics[width=0.99\textwidth]{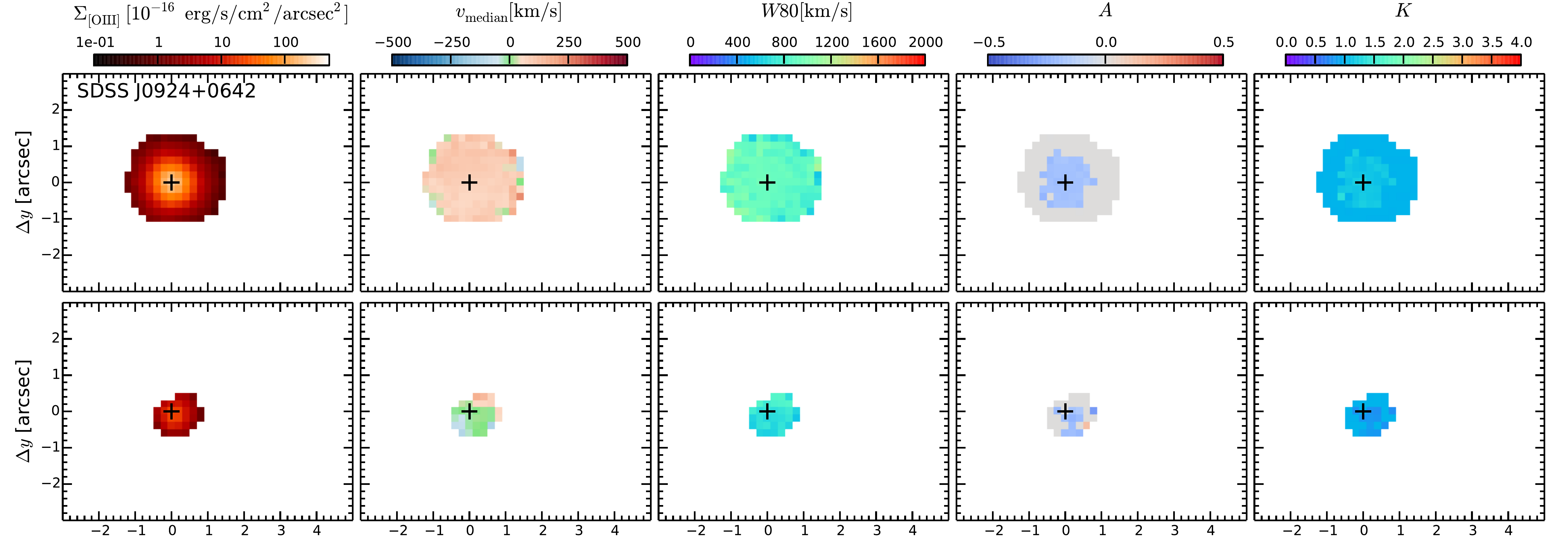}
\includegraphics[width=0.99\textwidth]{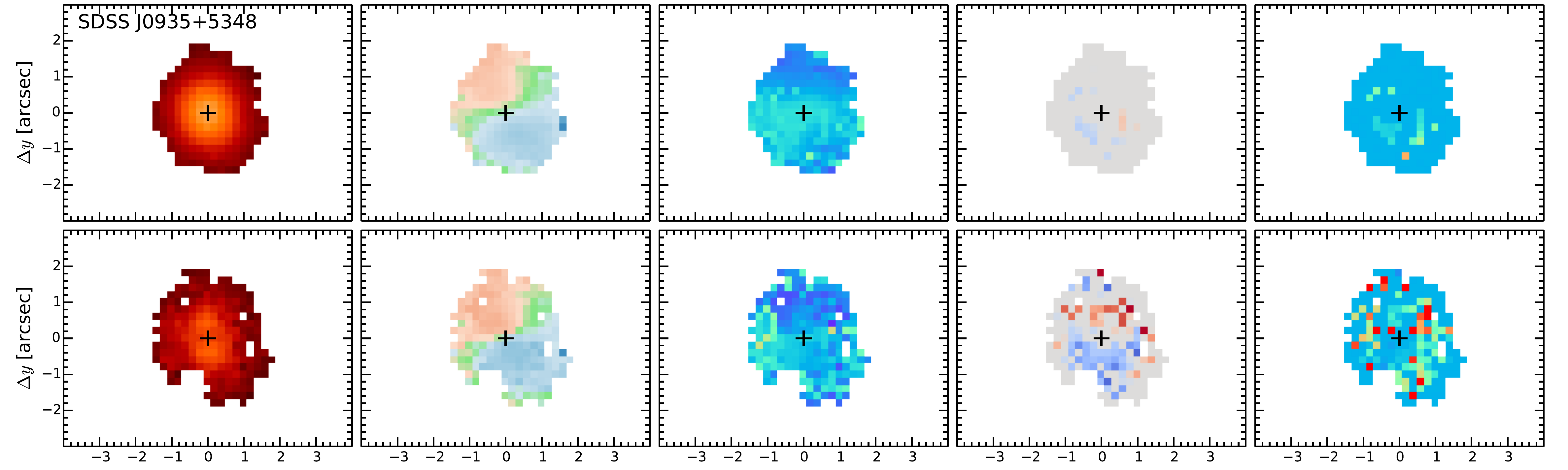}
\includegraphics[width=0.99\textwidth]{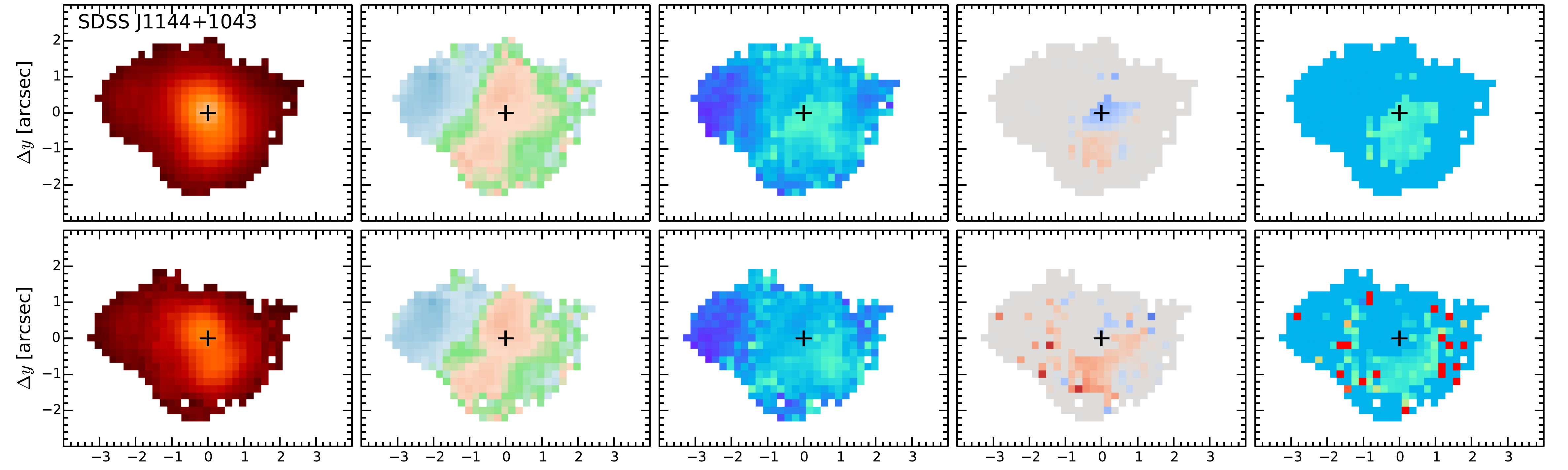}
\includegraphics[width=0.99\textwidth]{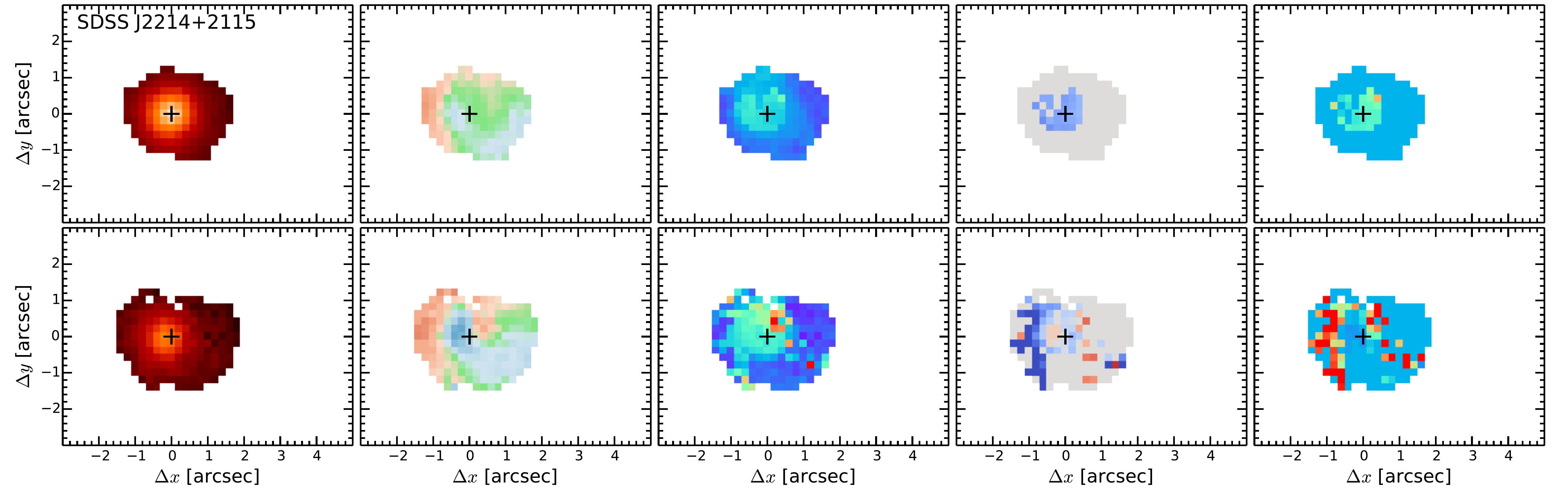}
\caption{continued.}
\end{figure*}

\section{Quantifying the impact of an unresolved NLR on ENLR measurements}\label{sect:comparison}
\begin{table*}
\caption{Basic parameters inferred for the ENLR before and after the deblending process.}
\label{tab:comparison}
\begin{footnotesize}
 \input{table2.tex}
 \end{footnotesize}
\end{table*}

\subsection{Comparison with the original measurements}
\begin{figure*}
\sidecaption
\includegraphics[width=12cm]{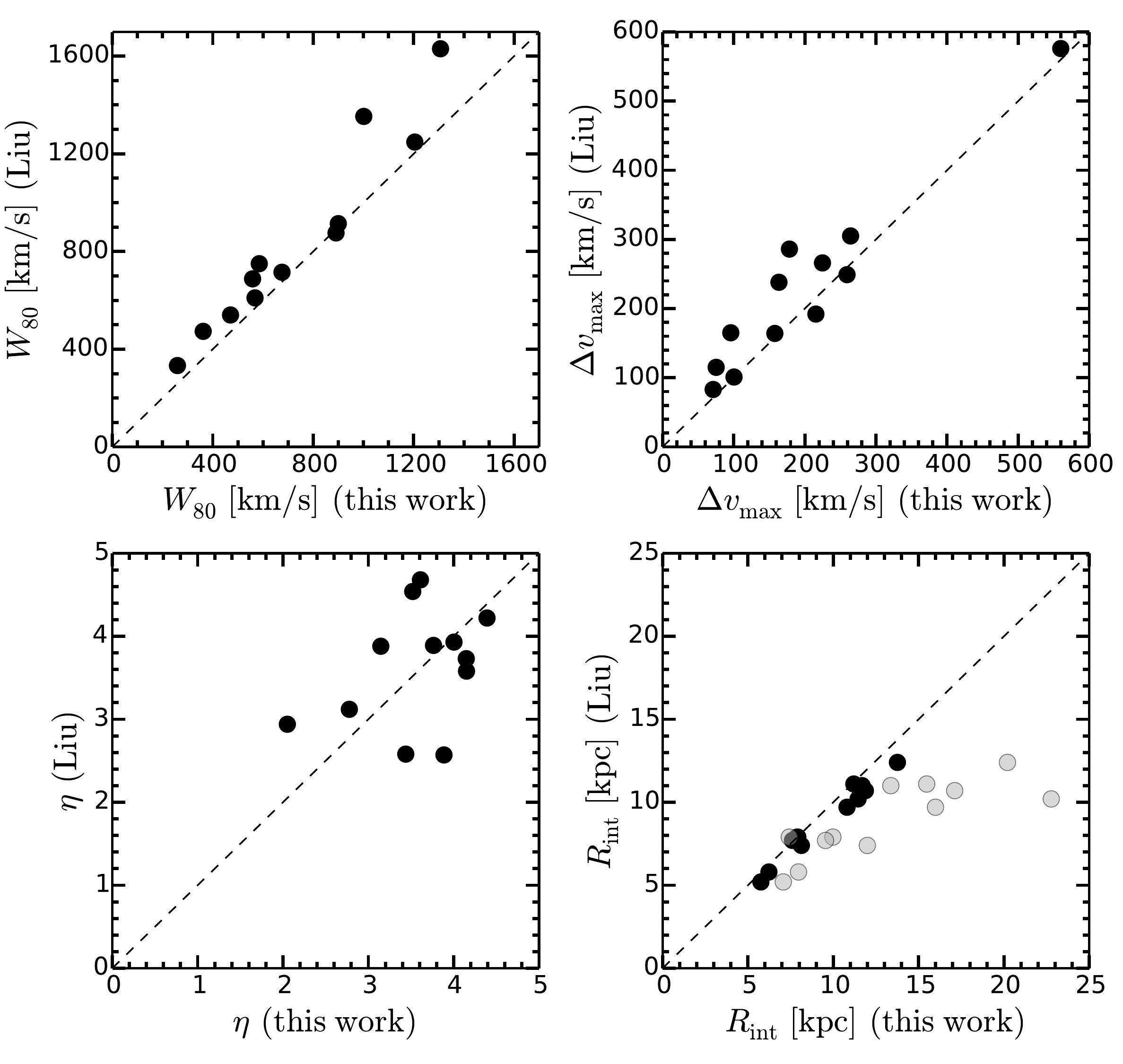}
\caption{Comparison of the total [\OIII] line width ($W_{80}$, upper left panel), maximum velocity range ($\Delta v$, upper right panel), size of the ENLR ($R_\mathrm{int}$, lower right panel) and 
the 
power-law slope of the total [\OIII] surface brightness profile ($\eta$, lower left panel) from the QSO as measured by \citet{Liu:2014} and our re-analysis. We find good agreement between 
measurements 
with only a weak systematic offset and a scatter consistent with the intrinsic accuracy of measurements. The surface brightness profile slope $\eta$ shows the greatest scatter because the actual 
range of the outer profile to measure $\eta$ is not clearly specified in \citet{Liu:2014}, so that our measurements are likely to not exactly match their methodology. For the ENLR we 
measure the isophotal radius at a surface brightness (corrected for cosmological dimming) of 
$\Sigma_{\mathrm{[\OIII]}}=10^{-15}\,\mathrm{erg}\,\mathrm{s}^{-1}\,\mathrm{cm}^{-2}\,\mathrm{arcsec}^{-2}$ 
(black circles) and the maximum projected distance to spaxels which exhibit the same threshold surface brightness locally (grey symbols).}
\label{fig:compare_Liu}
\end{figure*}

Since we performed a completely independent re-analysis, from the data reduction till the data analysis, compared to the work by \citet{Liu:2014}, we first want to test whether we recover their 
measurements if we follow the same methods as close as possible. In Fig.~\ref{fig:compare_Liu}, we show a comparison of the [\OIII] line width ($W_{80}$), maximum velocity range ($\Delta 
v$), the size of the ENLR ($R_\mathrm{int}$) and the power-law slope of the total [\OIII] surface brightness profile ($I_\mathrm{[\OIII]}(R)\sim R^{-\eta}$) over the range $1\arcsec-2.5\arcsec$ from 
the QSO as  measured by \citet{Liu:2014} and our own measurement from total [\OIII] line maps before deblending.

We find that our measurements are in good agreement with the values reported by \citet{Liu:2014}. Systematic difference are less than 20 per cent in all cases. Our measurements for $W_{80}$ 
is slightly smaller by $13\pm10$ per cent and also the maximum velocity range $\Delta v$ is smaller by $15\pm17$ per cent. For the latter, the rms is significant which is caused by the systematic 
uncertainties on the measurements of the radial velocities given that we have independently re-reduced the entire dataset. The most significant scatter is found for the outer radial [\OIII] surface 
brightness profile $\eta$. The reason for this is that the radius over which the slope is measured is not clearly defined in \citet{Liu:2014}. The arbitrary fitting range of 
$1\arcsec-2.5\arcsec$ along the major axis of the ENLR that we adopt here may simply not reflect the original prescription to measure this parameter. However, our mean value of $\langle \eta \rangle 
= 3.5\pm0.6$ is totally consistent with the measurements for the unobscured and obscured QSOs by \citet{Liu:2014}.

We can almost exactly reproduce the isophotal radius of the ENLR, based on the surface brightness of concentric annuli, with a rather small deviation of $5\pm5$ per cent. However, 
when the surface brightness of individual spaxels is concerned we can also define a ENLR size based on the largest projected distance from the QSO position to a single spaxel above 
the same threshold surface brightness. These ENLR sizes can be significantly larger than the azimuthally averaged isophotal radii reported by \citet{Liu:2014} and may explain the 
apparent flattening of the ENLR size -- QSO luminosity relation at high QSO luminosities \citep{Liu:2014,Hainline:2014}. It is beyond the scope of this paper to investigate this in detail.

In the following, we study how much the measurements change after separating the compact unresolved NLR from the ENLR. After the tests discussed above, any difference we find can be unambiguously 
attributed to the effect of beam smearing. Given the high signal-to-noise of the data we do not report uncertainties on measured quantities in Table~\ref{tab:comparison} which 
exhibits 
measurement errors of less than 0.1\,dex. Systematic uncertainties on the derived quantities will dominate the error budget by more than an order of magnitude so that we can safely ignore 
the measurement errors.

\begin{figure*}
\includegraphics[width=\textwidth]{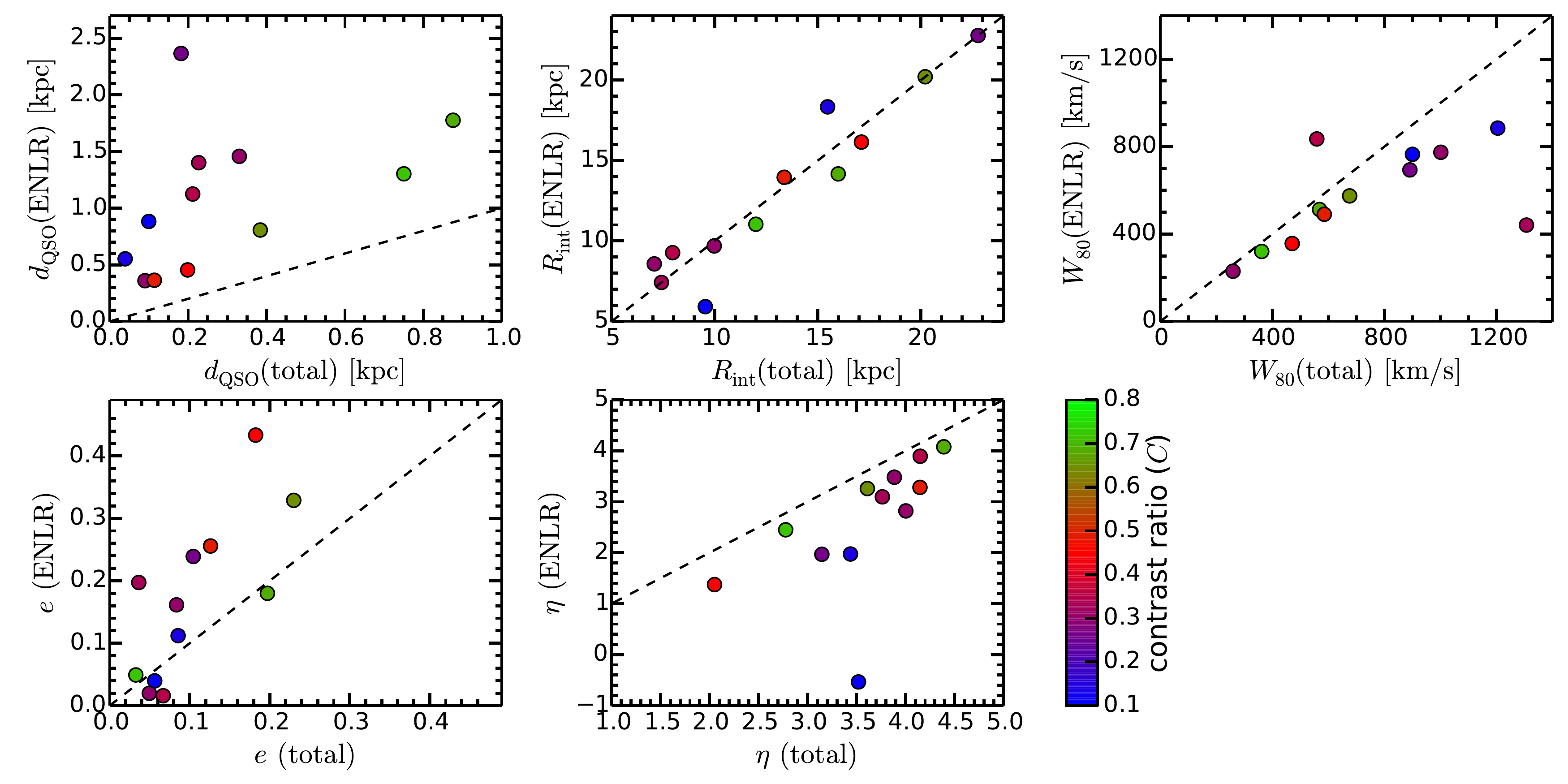}
\caption{Comparison of the total and ENLR [\OIII] line measurements for the centroid distance from the QSO ($d_\mathrm{QSO}$), maximum size out to a fixed surface brightness limit ($R_\mathrm{int}$), 
the median [\OIII] line width within the central $1''$ ($W_{80}$), the ellipticity of the central [\OIII] emitting region ($e$) and the outer power-law radial surface brightness slope ($\eta$). The 
1:1 relations are shown as a dashed line in each panel. The colour of each data point corresponds to the contrast ratio $C$ as indicated by the colour bar which is defined as the fraction of 
resolved to the total [\OIII] emission.}
\label{fig:comparison_deblending}
\end{figure*}

\subsection{Surface brightness distribution}
We observe some subtle but important  difference between the [\OIII] surface distribution for some objects, after subtracting the compact unresolved NLR contribution.  Here, we quantify the 
changes by means of a few important parameters. From the flux maps, we compute the total and ENLR [\OIII] flux ($f_\mathrm{[\OIII]}$) within the GMOS FoV from which we define a contrast ratio 
as $C=f_\mathrm{ENLR}/(f_\mathrm{NLR}+f_\mathrm{ENLR})$; the fraction of the ENLR to the total flux. In addition, we compute the flux-weighted centroid and ellipticity within a 
$2\arcsec\times2\arcsec$ sub-frame centred on the QSO position. From the flux-weighted centroid, we infer the apparent distance to the QSO position ($d_\mathrm{QSO}$), defined as the flux-weighted 
centre of the broad H$\beta$ distribution. All those measurements are summarized in Table~\ref{tab:comparison}.

We find that the contrast ratio $C$ spans a large range across the sample. In four QSOs the ENLR contributes more than 50 per cent to the total [\OIII] emission, whereas the ENLR contributes less 
than 10\,per cent in the most extreme case. As expected, the changes in the ENLR surface brightness distribution appear marginal if $C>0.5$.
Remarkably, the distance between the peak in the surface brightness distribution and the QSO position $d_\mathrm{QSO}$ is generally higher for pure unresolved emission (ENLR) than for the 
total light  as shown in Fig.~\ref{fig:comparison_deblending} (upper left panel).  The ratio between the offsets also increases with with decreasing contrast ratio $C$.
Those offsets of up to $\sim0.2''$ correspond to about $1-2$\,kpc at the redshift of the QSOs and imply that the ENLR is highly asymmetric. They appear much smaller in the 
total [\OIII] light which is clearly attributed to the bright unresolved NLR almost centred on the QSO. Only in two cases, SDSS~J0412$-$0511 and SDSS~J0753$+$3153, we can neither detect a large 
offset ($d<0.4$\,kpc) nor a strong elongation of the ENLR ($e\leq0.1$).

Something that is not strongly effected by the beam smearing is the size of the ENLR up to an intrinsic [\OIII] surface brightness threshold of 
$\Sigma_\mathrm{[\OIII]}>10^{-15}/(1+z)^{4}\,\mathrm{erg}\,\mathrm{s}^{-1}\,\mathrm{cm}^{-1}$, corrected for cosmological surface brightness dimming, which was defined in 
\citet{Liu:2013} and is an arbitrary choice. As we discussed before, 
we simply measure the distance to all the spaxels that are above the surface brightness threshold. Among those spaxels, the one with the greatest distance defines the ENLR size. Apparently, 
the ENLR dominates the emission at large radii sufficiently well, so that we find that no significant change of $R_\mathrm{int}$ as a function of contrast ratio (Fig.~\ref{fig:comparison_deblending}). 
This is consistent with the studies of \citet{Hainline:2013,Hainline:2014} who find that the ENLR maybe at most 0.1-0.2\,dex smaller after a full PSF convolution of the [\OIII] surface brightness 
distribution. Only for SDSS~J0924$+$0624 do we recover a substantially smaller ENLR size by 40\,per cent (0.4\,dex). Here, the the total emission is dominated by unresolved emission 
($C<0.2$) and is most strongly affected by the beam smearing. 

If we look at the power-law slope of the outer surface brightness profile between $1''-2.5''$ away from the QSO, we find that the slope becomes flatter with decreasing contrast ratio. 
Although we measure the slope outside the formal seeing disc, the wings of the PSF still contribute to the surface brightness beyond $1''$ making the profile steeper. At the lowest 
contrast ratio, SDSS~J0924$+$0624 stands out again, because the size of the ENLR is much smaller than $2.5''$ after subtracting the unresolved emission. In this case, the slope actually does not 
make sense 
at it is dominated by noise over most of the range. Therefore, we think that the power-law slope of the surface brightness is an ill-defined quantity if the beam smearing is not taken into account for 
data at the given spatial resolution.

\subsection{Spatially resolved kinematics}
In Table~\ref{tab:comparison}, we also report the characteristic ENLR [\OIII] line width $W_{80}$ as the median of all individual spaxel measurements within $<$0\farcs6 around the QSO position.
Three QSOs in the sample appear to show an [\OIII] line width of $W_{80}>1000\,\mathrm{km\,s}^{-1}$ on kpc scales consistent with \citet{Liu:2014}. We find that in all those cases, the unresolved NLR 
is at least as bright as the entire ENLR with $C<0.5$ and that the [\OIII] line width in the ENLR reduces significantly to $W_{80}<800\,\mathrm{km\,s}^{-1}$ after the deblending of the NLR. The 
most extreme difference between NLR and ENLR kinematics is observed for the QSO SDSS~J0304+0022 with $W_{80}\sim1500\,\mathrm{km\,s}^{-1}$ for the NLR and almost completely quiescent kinematics for 
the ENLR with $W_{80}\sim400\,\mathrm{km\,s}^{-1}$. The opposite happens for QSO SDSS~J2214+2115 for which we detect significantly broader lines in the ENLR after removing the NLR contribution. In all 
the other cases the line widths are either fully consistent with each other or slightly smaller by 100--200$\,\mathrm{km\,s}^{-1}$. 

Depending on the contrast ratio, we see more detailed structure in the velocity field after the QSO-host deblending. An extreme case is SDSS~J0924$+$0642 ($C<0.2$) where we see a 
symmetric velocity gradient across the nucleus with an amplitude of $\pm230\,\mathrm{km\,s}^{-1}$. The velocity field in the total light appears flat, because the velocity of the unresolved NLR 
dominates over a significant area due to the seeing. The signature for symmetric velocity gradients is also clearly enhanced in the case of SDSS~J0304$-$0707, SDSS~J0753+3153 and SDSS~J2214$+$2115. 
Whether those gradients are due to ordered rotation of a gas disc or indicate bipolar outflows is unclear at this point.

Another special case is SDSS~J0304$+$0022 for which we detect a huge offset of $>500\,\mathrm{km\,s}^{-1}$ in the radial velocity close to the QSO position after subtracting the unresolved emission. 
The [\OIII] line in this QSO has an exceptional broad blue-shifted component, but only the narrow [\OIII] component is actually spatially resolved. Thus, the radial velocity measured from  the total 
light is dominated by the unresolved emission up to the radius where the ENLR emission starts to dominate the [\OIII] line shape. 

\subsection{Spatially-resolved line ratios}
A key diagnostic for the ionization conditions of the ENLR is the [\OIII]/H$\beta$ line ratio. \citet{Liu:2013} measured the line ratio across the ENLR for their sample of obscured QSOs as a function 
of [\OIII] surface brightness and distance from the QSO. They reported an almost constant ratio [\OIII]/H$\beta\sim10$ up to a characteristic radius of $R\sim7$\,kpc after which the line ratio 
is dropping continuously. Here, we present the same analysis for the unobscured QSOs which was not presented in \citet{Liu:2014}. In Fig.~\ref{fig:line_ratios}, we show the [\OIII]/H$\beta$ line 
ratio for all the spaxels with a S/N$>$5 before and after subtracting the unresolved emission contribution. The imposed S/N condition leads to a detection limit for the line ratio that varies with 
the [\OIII] surface brightness given the fixed depth of the a given dataset.

\begin{figure*}
 \centering
 \includegraphics[width=0.97\textwidth]{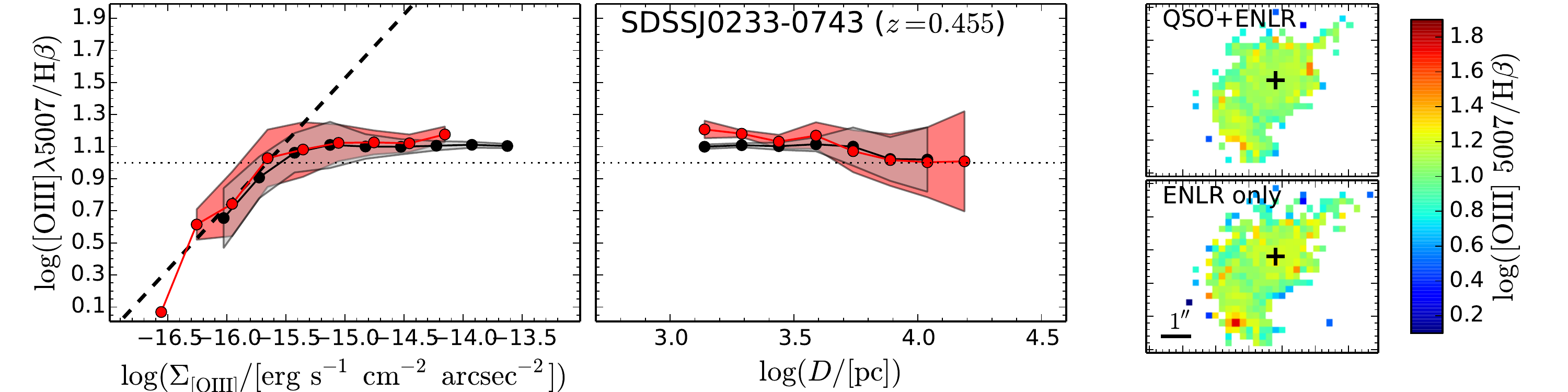}
 \includegraphics[width=0.97\textwidth]{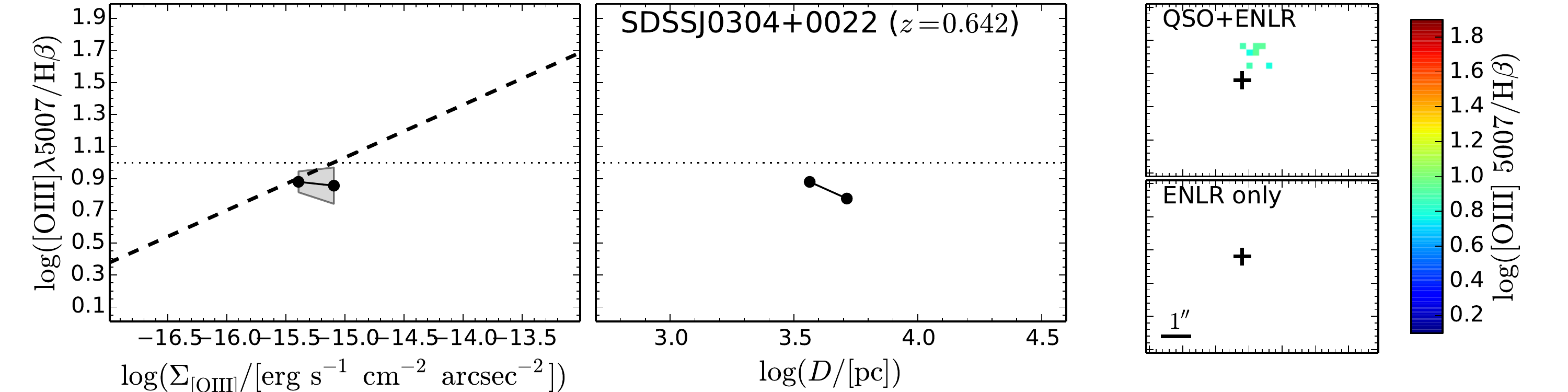}
 \includegraphics[width=0.97\textwidth]{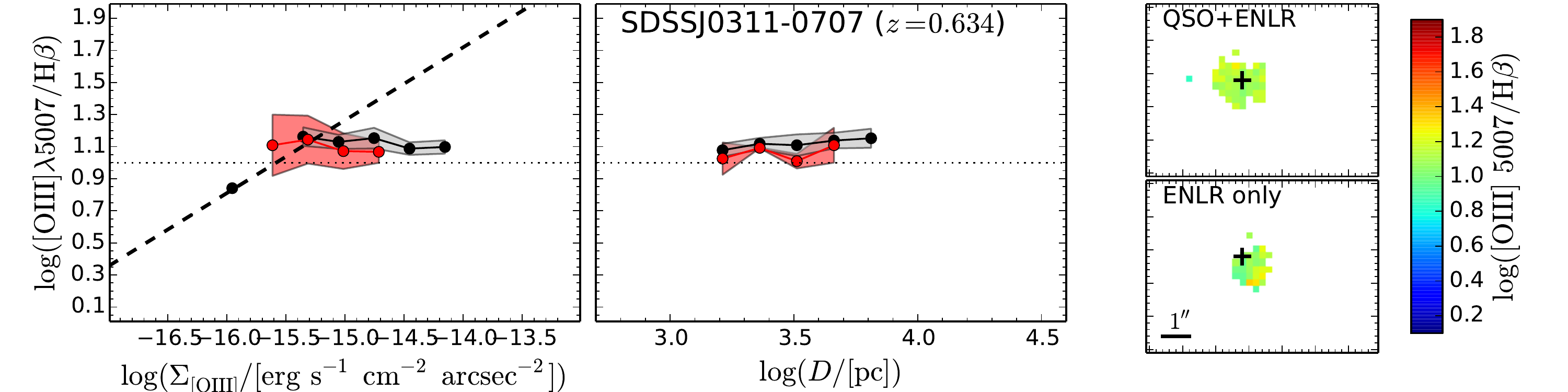}
 \includegraphics[width=0.97\textwidth]{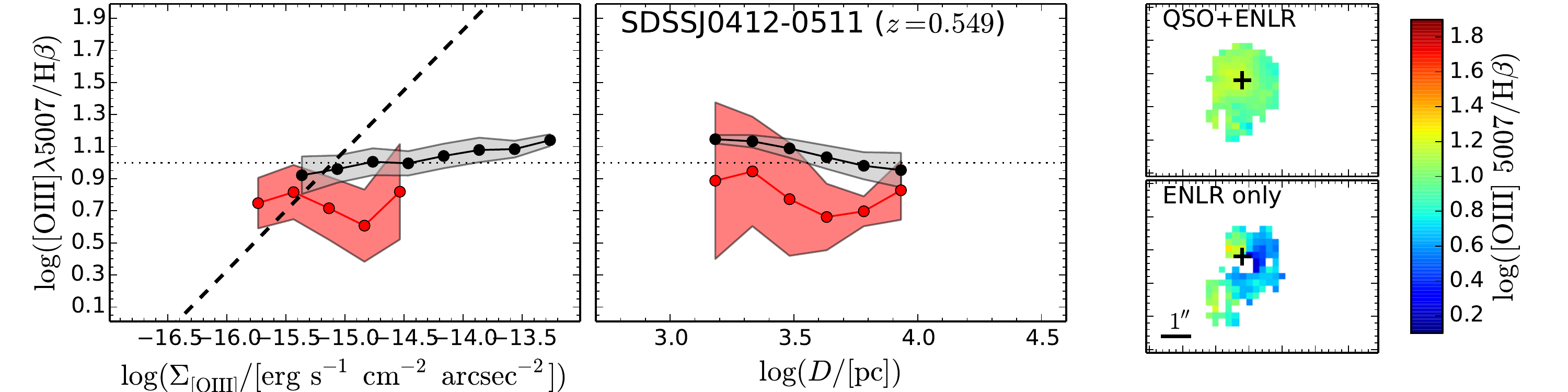}
 \includegraphics[width=0.97\textwidth]{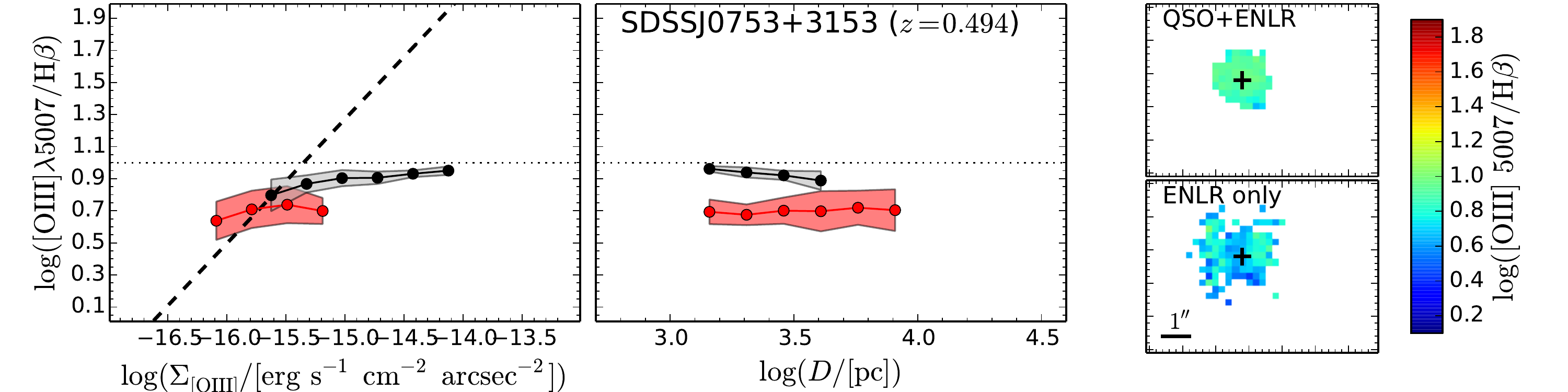}
 \caption{{\it Left panel:} [\OIII]/H$\beta$ emission-line ratio as a function of [\OIII] surface brightness for the original data after removing the broad H$\beta$ line (black data points) and 
after performing the QSO-host galaxy deblending (red data points). The shaded area indicates the rms of line ratios within a given bin. The dashed line indicates the $3\sigma$ detect limit for 
H$\beta$ for a given [\OIII]/H$\beta$ ratio given the estimated noise in the unblended data. {\it Middle panel:}  Line ratios as a function of distance 
$D$ from the QSOs. {\it Right panel:} Line ratio maps before and after the QSO-host galaxy deblending.}
 \label{fig:line_ratios}
\end{figure*}
\addtocounter{figure}{-1}

\begin{figure*}
 \centering
 \includegraphics[width=\textwidth]{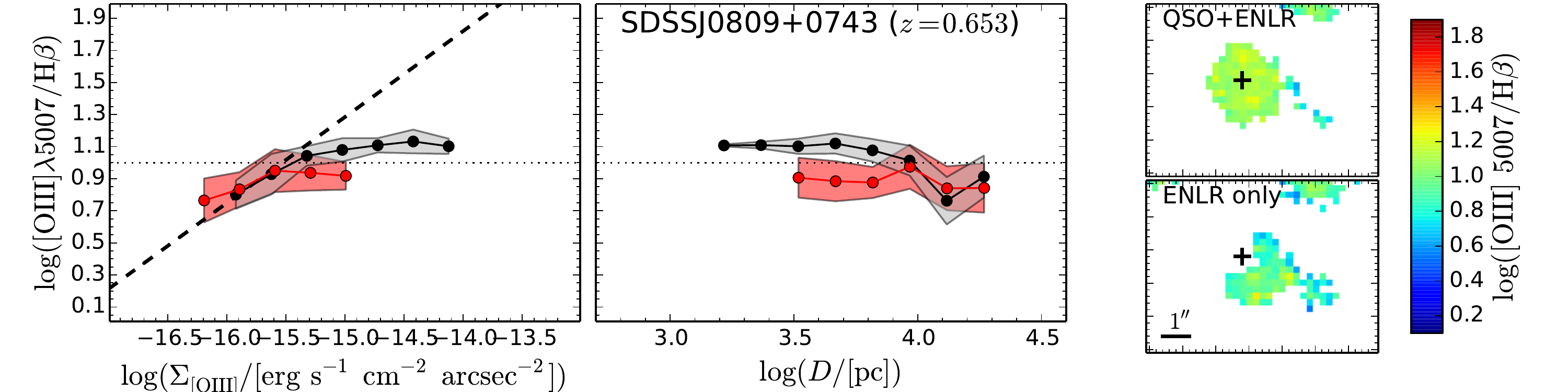}
 \includegraphics[width=\textwidth]{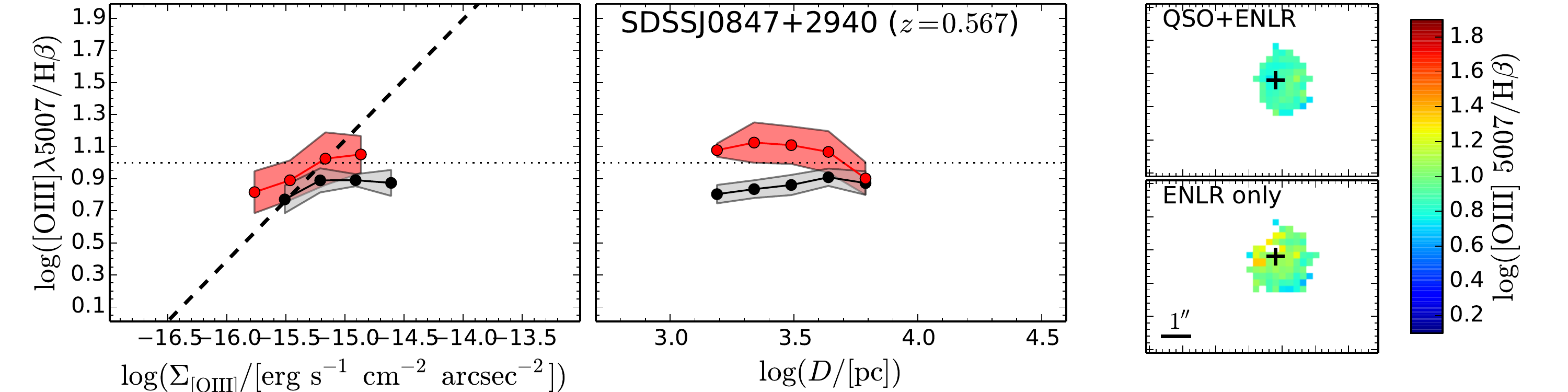}
 \includegraphics[width=\textwidth]{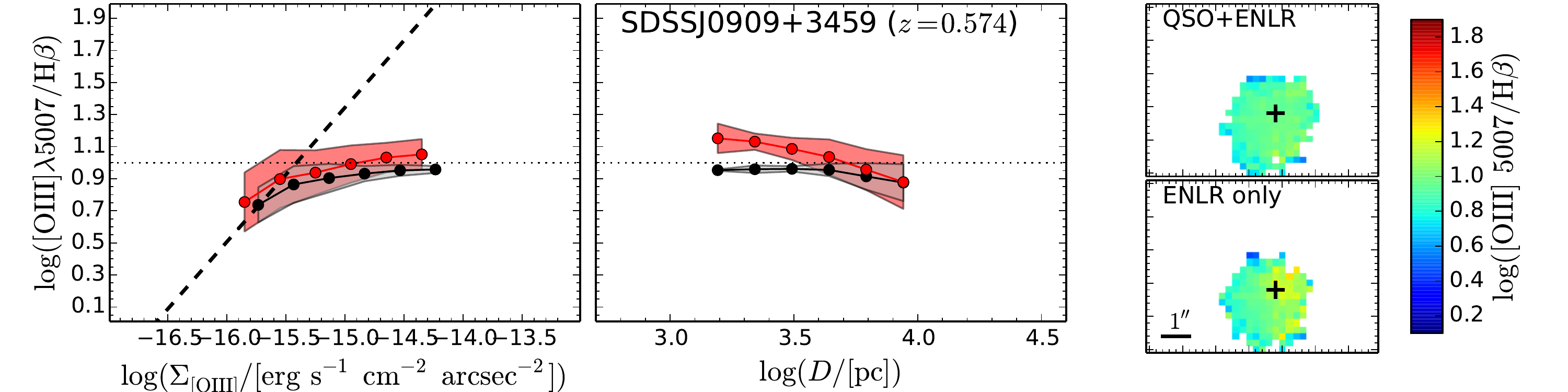}
 \includegraphics[width=\textwidth]{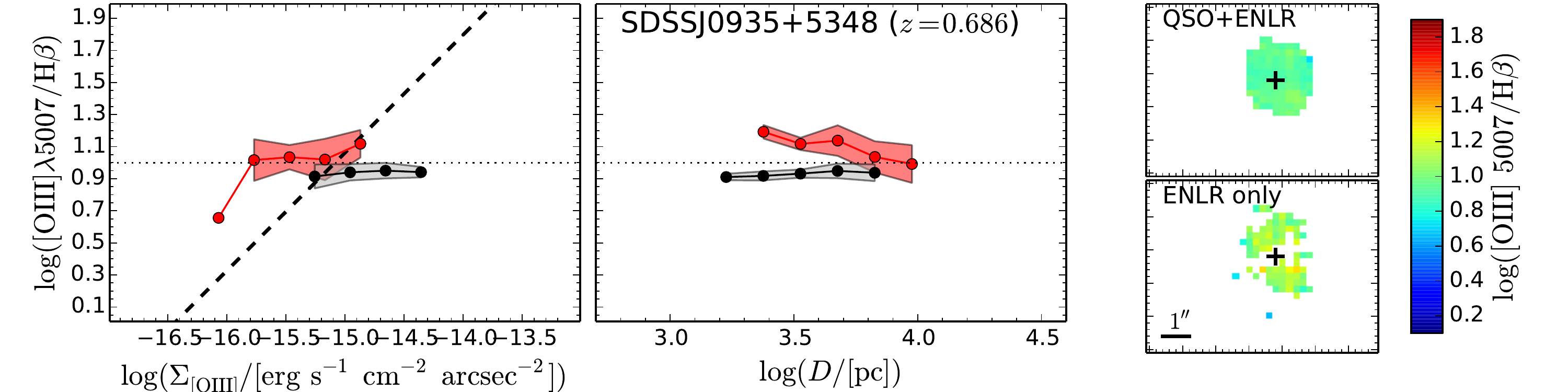}
 \includegraphics[width=\textwidth]{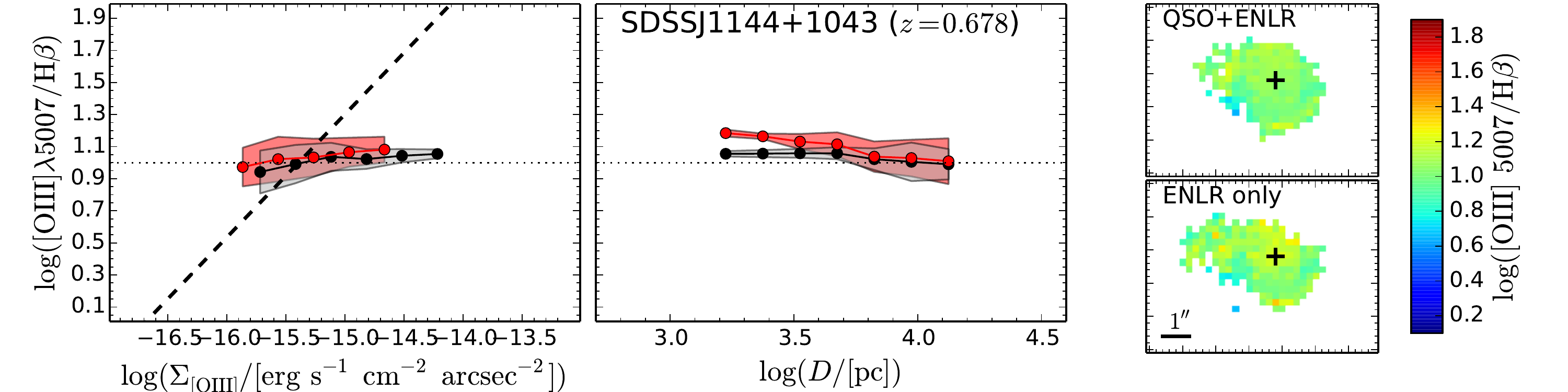}
 \caption{continued.}
\end{figure*}
\addtocounter{figure}{-1}
\begin{figure*}
 \centering
\includegraphics[width=\textwidth]{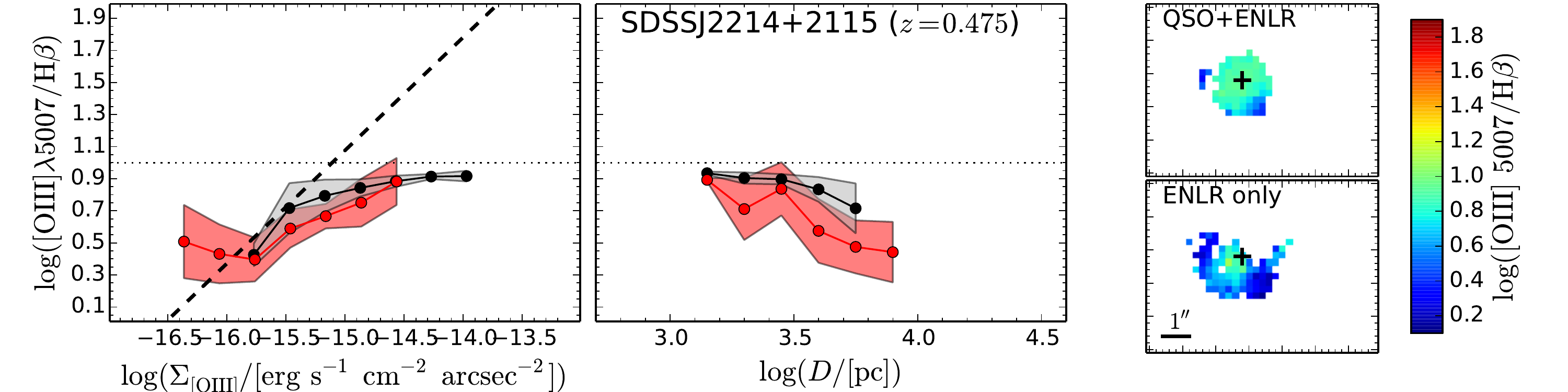}
\caption{continued.}
\end{figure*}
In contrast to  \citet{Liu:2013}, we do not detect a strong break in the line ratio, which remains flat close to $\log(\mathrm{[\OIII]/H}\beta)\sim1$ across the entire range of 
$\Sigma_{\mathrm{[\OIII]}}$ and $D$. The deblending of the NLR and ENLR does not change the line ratios in most cases and confirms that the ENLR is photoionized by the AGN out to large distances. 
Exceptions from the flat distributions are SDSS~J0233-0743 and SDSS~J2214+2115, which show a systematic decrease of $\mathrm{[\OIII]/H}\beta$ at low surface brightness. This decrease can be explained 
either by ionization of young stars in star forming regions, slow shocks $<500\,\mathrm{km\,s}^{-1}$ in the ISM or if photoionization by the AGN changes from the ``ionization``  to the 
''matter-bounded`` region. The latter scenario has been favoured by \citet{Liu:2013} as an explanation for the strong decrease in $\log(\mathrm{[\OIII]/H}\beta)$ after a well-defined break radius.

One important aspect to consider here are the actual detection limits for the lines. When $\Sigma_{\mathrm{[\OIII]}}$ decreases, the H$\beta$ line may already be below the detection limit depending 
on the intrinsic line ratio. Since only spaxels are considered for which \textit{both} lines are detected with $>3\sigma$ confidence, a bias is introduced  towards low [\OIII]/H$\beta$ line ratios 
with decreasing $\Sigma_{\mathrm{[\OIII]}}$ if low line ratios are present in the data. It is unclear at this point which role this effect plays in the analysis of the corresponding obscured QSOs 
sample \citep{Liu:2013,Liu:2013b}.

\section{AGN outflow energetics}\label{sect:feedback}
The estimation of the ionized gas outflow energetics and mass outflow rate is a difficult task and usually depends on assumption on parameters that are not directly constrained by the 
data. In particular, the lack of spatial resolution usually does not allow to directly constrain the geometry of the ionized gas outflows. This is even worse for high-redshift AGN where the spatial 
resolution is limited to a few kpc per resolution element. Here, we primarily focus on the comparison of estimates from different models and evaluate how strongly they are affected by contributions 
from an unresolved source due to beam smearing.

\subsection{Ionized gas mass and kinetic energy}
The amount of ionized gas is set by the amount of ionized hydrogen which can be estimated from the photons emitted by the recombination lines. Adopting ``Case B'' recombination for the low-density 
limit and a gas temperature of 10\,000\,K,  we expect an intrinsic Balmer line decrement of H$\alpha$/H$\beta=2.85$ \citep{Osterbrock:2006}. The ionized gas mass can then be approximated from the 
H$\beta$ luminosity following 
\citep{Osterbrock:2006} as
\begin{equation}
M_\mathrm{ion}=\frac{1.4m_\mathrm{p}}{n_\mathrm{e}\alpha_{\mathrm{H}\beta}^\mathrm{eff}h\nu_{\mathrm{H}\beta}}L_{\mathrm{H}\beta}=10^7\left(\frac{100\mathrm{cm}^{-2}}{n_\mathrm{e}}\right)\left(\frac{
L _{\mathrm{H}\beta}}{10^{41}\mathrm{erg\,s}^{-1}} \right)\mathrm{M}_\odot  \label{eq:mass}
\end{equation}
where $m_\mathrm{p}$ is the proton mass, $n_\mathrm{e}$ is the electron density and $h$ is the Planck constant. Although H$\beta$ is covered in the observed wavelength range, it suffers from a 
much lower S/N per resolution element. We therefore use the bright [\OIII] as a surrogate for H$\beta$ with a  line ratio of [\OIII]/H$\beta\sim10$. Adopting this fixed line ratios is accurate within 
$\pm0.2$\,dex for all objects as verified by the line ratio distribution (Fig.~\ref{fig:line_ratios}).  In fact this provides a lower limit for the ionized gas mass as we do not apply any correction 
for internal dust extinction.   

The greatest uncertainty in our case is the unconstrained electron density $n_\mathrm{e}$. We cannot infer it from the data itself because no density-sensitive lines are in the covered wavelength 
range. For the following calculations, we adopt an electron density of $n_\mathrm{e}\sim 100\,\mathrm{cm}^{-2}$ as a reference. This value is a typical value observed in the ENLR around luminous QSOs 
\citep[e.g.][]{Husemann:2016a}. However, the density has a large range since it is decreasing with distance \citep[e.g.][]{Bennert:2006a,Bennert:2006b} from $n_\mathrm{e}\sim 
1000\,\mathrm{cm}^{-2}$ in the NLR on $100$\,pc scales \citep[e.g.][]{Vaona:2012} and $n_\mathrm{e}\sim 10\,\mathrm{cm}^{-2}$ in the very extended and diffuse medium on kpc scale 
\citep[e.g.][]{Liu:2013b}.

The total kinetic energy of the ionized nebulae is split into bulk motion $v_\mathrm{g}$ and the turbulent motion  $\sigma_g$ of the gas as measured from the emission lines. With the assumption of 
constant electron density we can simply integrate the localized kinetic energy per spatial pixel at position $x$ and $y$ leading to
\begin{equation}
  E_\mathrm{kin} = \frac{1}{2}\sum_{x,y}M_{\mathrm{ion}}(x,y)\left(v_\mathrm{g}(x,y)^2+\sigma_\mathrm{g}(x,y)^2\right)
\end{equation}

The results for the estimated ionized gas mass and the kinetic energy are listed in Table~\ref{tbl:energetics}. The total ionized gas mass is in the range of 0.6--$10\times 10^8\mathrm{M}_\odot$ 
with a mean of $2\times 10^8\mathrm{M}_\odot$. Although \citet{Liu:2014} did not estimate the ionized gas mass for this unobscured QSO sample, they reported a similar ionized gas 
mass of $6\times 10^8\mathrm{M}_\odot$ for their obscured QSO sample \citep{Liu:2013b} following the same assumptions.

Here, we derive the kinetic energy from the kinematics of the total and the ENLR [\OIII] line profile distribution across the field separately. We note that in the majority of cases the kinematic 
energies do not change significantly.  Only QSOs with a lower contrast ratio $C$ show a clear difference which is 
caused by a lower ionized gas mass in the ENLR and a lower line width on kpc scales which is reducing the turbulent energy term. Wee obtain a mean kinetic energy of $10\times10^{55}$\,erg and 
$6\times10^{55}$\,erg, for the total and the ENLR energy, respectively. This is about 1\,dex lower than reported by \citet{Liu:2013b} for the unobscured QSOs, because they assumed 
a constant outflow velocity of $760\,\mathrm{km\,s}^{-1}$ across the entire nebulae. 

Assuming a time scale $\tau$ for the kinetic energy injection one can roughly estimate a kinetic power $\dot E_\mathrm{kin}=E_\mathrm{kin}/\tau$. Usually a time scale of about $10^7$\,yr is assumed 
for the life time of a luminous QSO phase, but \citet{Schawinski:2015} recently suggested a much shorter time scale of $10^5$\,yr.  This yields kinetic powers of 
$2\times10^{41}\,\mathrm{erg\,s}^{-1}$ and 
$2\times10^{43}\,\mathrm{erg\,s}^{-1}$, respectively, for the two time scales.

As argued in many studies \citep[e.g.][]{Nesvadba:2006,Cicone:2014}, these estimates are lower limits to the actual kinetic energy, because the geometry and projection effect are not 
taken into account. Therefore, models for the outflow have been used to improve the estimates. Below we describe the results we obtain for a spherical symmetric outflow \citet{Liu:2013b} and a 
conical outflow model \citep[e.g.,][]{CanoDiaz:2012,Cresci:2015}.

\begin{table*}
 \caption{Derived outflow energetics for different models}
 \label{tbl:energetics}
 \input{table3.tex}
\end{table*}

\subsection{Kinetic power and mass outflow rate}
\subsubsection{Spherical symmetric outflow model}
\citet{Liu:2013b} adopted a spherical and symmetric outflow geometry to estimate the energetics and outflow rate for the ionized gas around luminous obscured QSOs. They argued that a spherical 
geometry is strongly supported by the apparently round ENLR with almost constant broad lines out to kpc distances. They define a shell at distance $D$ through which they estimate the 
current kinetic power ($\dot E_\mathrm{kin}$) and mass outflow rate $\dot M$. They set $D$ to be radius at which they observed a break in the [\OIII]/H$\beta$ line ratio as a function of radius which 
they associate with the transition from ionization- to matter-bounded clouds, which occurs at $D\sim7$\,kpc for their sample of QSOs. Based on the assumptions of matter-bounded and 
pressure-confined clouds, spherical symmetry and  ionization equilibrium, \citet{Liu:2013} derived the following relations
\begin{multline}
 \frac{\dot 
E_\mathrm{kin}(D)}{2.6\times 10^{40}\mathrm{erg}\,\mathrm{s}^{-1}}=\left(\frac{\Sigma_{\mathrm{H}\beta}(D)}{10^{-15}\,\mathrm{erg\,s}^{-1}\mathrm{cm}^{-2}\mathrm{arcsec}^{-1}}\right)\\
\times \left(\frac{100\mathrm{cm}^{-3}}{n_\mathrm{e}}\right)\left(\frac{
v_\mathrm{out}}{100\mathrm{km\,s}^{-1}} \right)^3\left(\frac{\mathrm{kpc}}{D}\right)
\end{multline}
where $\Sigma_{\mathrm{H}\beta}(D)$ is the H$\beta$ surface brightness, corrected for the surface brightness dimming with redshift, at distance $D$ from the QSO, $v_\mathrm{out}$ 
is the outflow velocity and $n_e$ is the electron density. The corresponding mass outflow is then defined as $\dot M=2\dot E_\mathrm{kin}/v_\mathrm{out}^2$ which corresponds to
\begin{multline}
\frac{\dot 
M_\mathrm{out}(D)}{0.08\,\mathrm{M}_\odot\mathrm{yr}^{-1}}=\left(\frac{\Sigma_{\mathrm{H}\beta}(D)}{10^{-15}\,\mathrm{erg\,s}^{-1}\mathrm{cm}^{-2}\mathrm{arcsec}^{-1}}\right)\\
\times \left(\frac{100\mathrm{cm}^{-3}}{n_\mathrm{e}}\right)\left(\frac{
v_\mathrm{out}}{100\mathrm{km\,s}^{-1}} \right)\left(\frac{\mathrm{kpc}}{D}\right)
 \end{multline}
Since the maximum ENLR size drops to $R_\mathrm{max}=6\,$kpc after subtracting the unresolved emission for SDSS~J0924+0642 and no clear break radius in the line ratios is detected for any 
of the objects, we adopt a fixed radius of $D=5$\,kpc for which we compute the energetics. In this way we can consistently measure the mean [\OIII] surface brightness within 
$4\,\mathrm{kpc}<R<6\,\mathrm{kpc}$ for all objects and therefore achieve comparable estimates among the sample considering the similar luminosity of all QSOs. Given the low spatial resolution of the 
data there is not objective criterion to adjust the radius on an object-by-object basis for a comparative study.

In the spherical symmetric outflow model, \citet{Liu:2013b} predicted the line shape to vary across the field as a function of the distance from the QSO and measured radial velocity 
$v_z$ and adopted a power-law function for the radial luminosity distribution in [\OIII] with slope $\alpha = \eta +1$,
\begin{equation}
    I (D,v_z) \propto (1-(v_z/v_\mathrm{out})^2)^{0.5(\alpha-3)}D^{(1-\alpha)}
\end{equation}

Such a line shape parametrization implies that $W_{80} \sim 1.3 \times v_\mathrm{out}$ for a power-law slope $\eta\sim3.5$ and $W_{80} \sim 1.5 \times v_\mathrm{out}$ for a power-law slope 
$\eta\sim2.6$ which are the mean slopes for the total and ENLR radial profiles, respectively. In Table~\ref{tbl:energetics} we report the computed kinetic powers and mass outflow rates based 
on the prescription above and adopting an electron density of $n_\mathrm{e}=100\,\mathrm{cm}^{-3}$. We find a mean kinetic power of $\dot 
E_\mathrm{kin}(D=5\,\mathrm{kpc})=10^{45}\,\mathrm{erg\,s}^{-1}$ and mass outflow rate of $\dot M_\mathrm{out}(D=5\,\mathrm{kpc})=450\,M_\sun\,\mathrm{yr}^{-1}$ for the initial values and $\dot 
E_\mathrm{kin}(D=5\,\mathrm{kpc})=6\times 10^{43}\,\mathrm{erg\,s}^{-1}$ and mass outflow rate of $\dot M_\mathrm{out}(D=5\,\mathrm{kpc})=100\,M_\sun\,\mathrm{yr}^{-1}$ for the ENLR after subtracting 
the unresolved emission contribution, respectively. Thus, the difference is more than an order of magnitude for the kinetic power and a factor of four in the mass outflow rate, but strongly depends on 
the contrast ratio for each individual source. In the extreme case, the kinetic power drops by more than 2\,dex and 1\,dex in the mass outflow rate. 

The changes in the energetics we state above are only valid for the kinetic power and mass outflow rate going through a sphere at a distance of $D=5
$\,kpc. The lower rates are expected due to subtraction of the point-like component which leads to a lower mass and smaller outflow velocity at that radius. However, the decomposition into
unresolved and resolved emission also implies that outflow power and mass outflow rate may change with time/distance from the nucleus in particular on scales smaller than 1\,kpc. Therefore, it 
would be important to estimate also the outflow power in the unresolved component. By design this is impossible for this specific spherical outflow model as it requires to compute 
the emission-line surface brightness at a given radius. We therefore explore this difference between the NLR and ENLR energetics details based on a simple bi-conical 
outflow model below.

\subsubsection{Simple conical outflow model}\label{results:conical}
In nearby Seyfert galaxies, the NLR and associated outflows were often reported to have a (bi-)conical geometry 
\citep[e.g.][]{Mulchaey:1996,Schmitt:2003a,Crenshaw:2010,Mueller-Sanchez:2011}. The frequency of conical outflows, however, is still a matter of debate. In a recent study 
\citet{Fischer:2013} could detect conical outflows only in $\sim$1/3 of nearby AGN based on \emph{HST} long-slit spectroscopy. It is not clear at this point if this is due to misaligned slits, 
weak/small outflows or a different geometry. On one hand, a conical outflow geometry is a natural outcome of the unified AGN model that can be 
easily resolved and confirmed for many nearby AGN with HST, if present. On the other hand, the opening angle is expected to increase with AGN luminosity so that for luminous QSOs a quasi-spherical 
outflow cannot be ruled out.

The spatial resolution of these luminous AGN at higher redshift does not allow to directly constrain the 
geometrical parameters for this model. \citet{CanoDiaz:2012} and \citet{Cresci:2015} preferred a conical outflow geometry and adopted a simple model for their high-$z$ QSOs. The authors assumed a 
conical geometry with opening angle $\Omega$, uniformly distributed clouds with the same density and a constant outflow velocity. With the assumption of constant density clouds, the kinetic power and 
mass outflow rate become independent of the opening angle and the filling factor of the clouds within the cone and one derives the following relations:
\begin{eqnarray}
 \dot E_\mathrm{kin}(D) & = &\frac{3}{2}\frac{M_\mathrm{ion}v_\mathrm{out}^3}{D}\\
 \dot M_\mathrm{out}(D)  & = & 3\frac{M_\mathrm{ion}v_\mathrm{out}}{D} 
\end{eqnarray}
Assuming Case B recombination with an electron temperature of $T\sim10^4$\,K, we can replace $M_\mathrm{ion}$ with Eq.~\ref{eq:mass} which leads to
\begin{eqnarray}
\frac{\dot E_\mathrm{kin}(D)}{10^{40}\mathrm{erg}\,\mathrm{s}^{-1}} = \left(\frac{100\mathrm{cm}^{-3}}{n_\mathrm{e}}\right)\left(\frac{
L _{\mathrm{H}\beta}}{10^{41}\mathrm{erg\,s}^{-1}} \right)\left(\frac{
v_\mathrm{out}}{100\mathrm{km\,s}^{-1}} \right)^3\left(\frac{\mathrm{kpc}}{D}\right)\\
  \frac{\dot M_\mathrm{out}(D)}{3\mathrm{M}_\odot\,\mathrm{yr}^{-1}} = \left(\frac{100\mathrm{cm}^{-3}}{n_\mathrm{e}}\right)\left(\frac{
L _{\mathrm{H}\beta}}{10^{41}\mathrm{erg\,s}^{-1}} \right)\left(\frac{
v_\mathrm{out}}{100\mathrm{km\,s}^{-1}} \right)\left(\frac{\mathrm{kpc}}{D}\right)
\end{eqnarray}
To be consistent with the estimates based on the spherical outflow model, we measure the [\OIII] luminosity within $D=5$\,kpc converted to H$\beta$ luminosity with a factor of 0.1 and assume 
again an electron density of $n_\mathrm{e}=100\,\mathrm{cm}^{-3}$. The outflow velocities are assumed to be maximum velocities in the works of \citet{CanoDiaz:2012} and \citet{Cresci:2015} so that we 
assume $W_{80}$ to be the representative outflow velocity. In this case, we obtain kinetic powers and mass outflow rates as reported in Table~\ref{tbl:energetics} with a mean kinetic 
power of $\dot E_\mathrm{kin}(D=5\mathrm{kpc})=3\times 10^{43}\,\mathrm{erg\,s}^{-1}$ and mass outflow rate of $\dot M_\mathrm{out}(D=5\mathrm{kpc})=85\,M_\sun\,\mathrm{yr}^{-1}$ for the initial 
values and $\dot E_\mathrm{kin}(D=5\mathrm{kpc})=2\times 10^{42}\,\mathrm{erg\,s}^{-1}$ and mass outflow rate of $\dot M_\mathrm{out}(D=5\mathrm{kpc})=16\,M_\sun\,\mathrm{yr}^{-1}$ for the ENLR after 
subtracting the  contribution of unresolved emission, respectively. The change due to the beam smearing of the unresolved NLR is again about 1\,dex for the kinetic power and a factor of five in mass 
outflow rate. Furthermore, the mean kinetic power and outflow rate is almost two orders of magnitude lower in the conical compared to the spherical outflow model.

In the same way, we can roughly estimate the kinetic power and mass outflow rate at smaller distances from the unresolved NLR component. Consequently, we adopt the assumptions for 
the unresolved NLR of a cone with a size of 1\,kpc, a slightly higher density of $n_e\sim600\,\mathrm{cm}^{-3}$ and a corresponding outflow velocity based on the [\OIII] line width in the 
unresolved QSO spectrum. Given that the assumed simple conical model is independent of the opening angle by design, we also consider it valid for an approximation for a spherical 
model given that we cannot constrain the outflow geometry for the unresolved NLR. Under these assumptions we obtain up to an order of magnitude higher kinetic power and outflow rates in the compact 
NLR than in the ENLR (see Table~\ref{tbl:energetics}). While this may indicate much more powerful outflows close to the nucleus these values have to be taken with a grain of salt. The 
assumption of constant-density clouds across the cone is expected to be strongly violated on these small scales and can vary by orders of magnitude up to several $1000\,\mathrm{cm}^{-3}$. Without 
spatially resolving the electron density via density-sensitive emission-lines, like done for local Seyfert galaxies with HST \citep[e.g.][]{Crenshaw:2000,Rice:2006,Storchi-Bergmann:2010,Fischer:2013}, 
 no firm conclusions can be made. However, it is certainly possible that a high mass outflow rate is still confined to a region less than 1\,kpc and has not travelled throughout the host galaxy yet.

\section{Discussion}\label{sect:discussion}
\subsection{Morphology and incidence of kpc scale outflows}
A major result of the work by \citet{Liu:2013b} and \citet{Liu:2014} is that the ENLR appears round with a constant line width of $W_{80}\sim1000\,\mathrm{km\,s}^{-1}$ on kpc scales around luminous 
QSOs at redshift $z\sim0.6$ irrespective of their type. This notion would naturally imply large opening angles for the escape of the AGN radiation out to large distance and therefore favour an 
almost spherical outflow geometry. This is in stark contrast to the results of \citet{Husemann:2013a} who reported rather elongated or even one-sided shapes of the ENLR around unobscured QSOs at 
$z<0.3$. One big difference in the analysis of \citet{Husemann:2013a} is the deblending of an unresolved NLR and the ENLR based on the PSF reconstructed from the broad 
H$\beta$ emission line. 

Here, we have resolved this issue and performed a consistent analysis of the unobscured QSO sample of \citet{Liu:2014} with the same deblending technique used by \citet{Husemann:2013a} to 
allow a fair comparison. In many cases, we recover one-sided or elongated structures in the ENLR close to the nucleus that was previously hidden underneath. The dominant emission of a bright 
unresolved NLR necessarily produces round structure caused by the seeing. In addition, the obtained offsets in the peak intensity of the ENLR about 1-2\,kpc are a striking feature and implies that 
the ionization has a preferred direction which disfavours a wide-angle ionization/outflow scenario on kpc scales. Nevertheless, we cannot rule out 
that the outflow associated with the compact unresolved NLR has a very different geometry, spherical or something completely different, and that only the radiation field on large 
scales appears conical or asymmetric.

\citet{Hainline:2013,Hainline:2014}  carefully considered the beam smearing effect on the ENLR in deep long-slit observations for a large sample of obscured QSOs. They 
reported a small overestimation of the ENLR size by 0.1--0.2\,dex if the beam smearing by the seeing is not considered. This is also consistent with the results of our analysis except for 
SDSS~J0924+0642 where the difference reaches even 0.3\,dex due to a very low contrast ratio. Hainline et al. assumed a Sersi\'c or Voigt profile for the ENLR light distribution, but 
enforcing azimuthal symmetry. Thus, they are not sensitive to asymmetries in the [\OIII] light distribution that can only be mapped using 3D spectroscopy or narrow-band imaging. Furthermore, with just 
a single slit it is impossible to measure the elongation of the ENLR and a reliable maximum extension. Very extended emission on $>$20\,kpc scales can be easily missed, as in in case of 
SDSS~J0809+0743, if the slit is not aligned with these structures.

It is well known that the forbidden lines, in particular the [\OIII] lines, are systematically asymmetric with a blue wing caused by a broad and blue-shifted emission line component 
\citep[e.g.][]{Heckman:1981,Whittle:1985a,Mullaney:2013,Shen:2015, Woo:2016}. 
This is usually interpreted as a signature of a fast bipolar AGN outflow were the receding side is obscured by the dust screen from the host galaxy. There is an increasing number of studies which 
report broad emission lines with line widths of $1000\,\mathrm{km\,s}^{-1}$ on kpc scale in luminous QSO at low \citep{Greene:2011,Liu:2013b,Harrison:2014,Liu:2014,McElroy:2015} and high redshift 
\citep{CanoDiaz:2012,Carniani:2015,Cresci:2015,Brusa:2015,Perna:2015}. However, all those studies lack a proper discussion of the impact of beam smearing on the observed 
light intensity profile of the broad [\OIII] line component to verify the real size of the outflows. 

Based on our re-analysis of the unobscured QSO sample of \citet{Liu:2014}, we find that the line width of [\OIII] is significantly broader in the 
unresolved NLR than in the ENLR on kpc scales. In particular the cases where the [\OIII] line appears broader than $1000\,\mathrm{km\,s}^{-1}$ (FWHM) on kpc scales reduces by several 
$100\,\mathrm{km\,s}^{-1}$ down to $400\,\mathrm{km\,s}^{-1}$ in the most 
extreme case of SDSS~J0304+0022 when the spatially unresolved component is removed. From the twelve QSOs in the \citet{Liu:2014} sample, eight have a line width $W_{80}>900\,\mathrm{km\,s}^{-1}$ in 
the NLR of which only three QSOs show a line width of $\sim$800$\,\mathrm{km\,s}^{-1}$ on kpc scales. There is only one case, SDSS~J2214+2115, where we find an increase in the line width on kpc 
scales which indicates a very powerful extended outflow. However, in general our analysis strongly favours outflows that slow down as they expand from the $\ll$1\,kpc scales of the NLR. 

\begin{figure*}
\includegraphics[width=\textwidth]{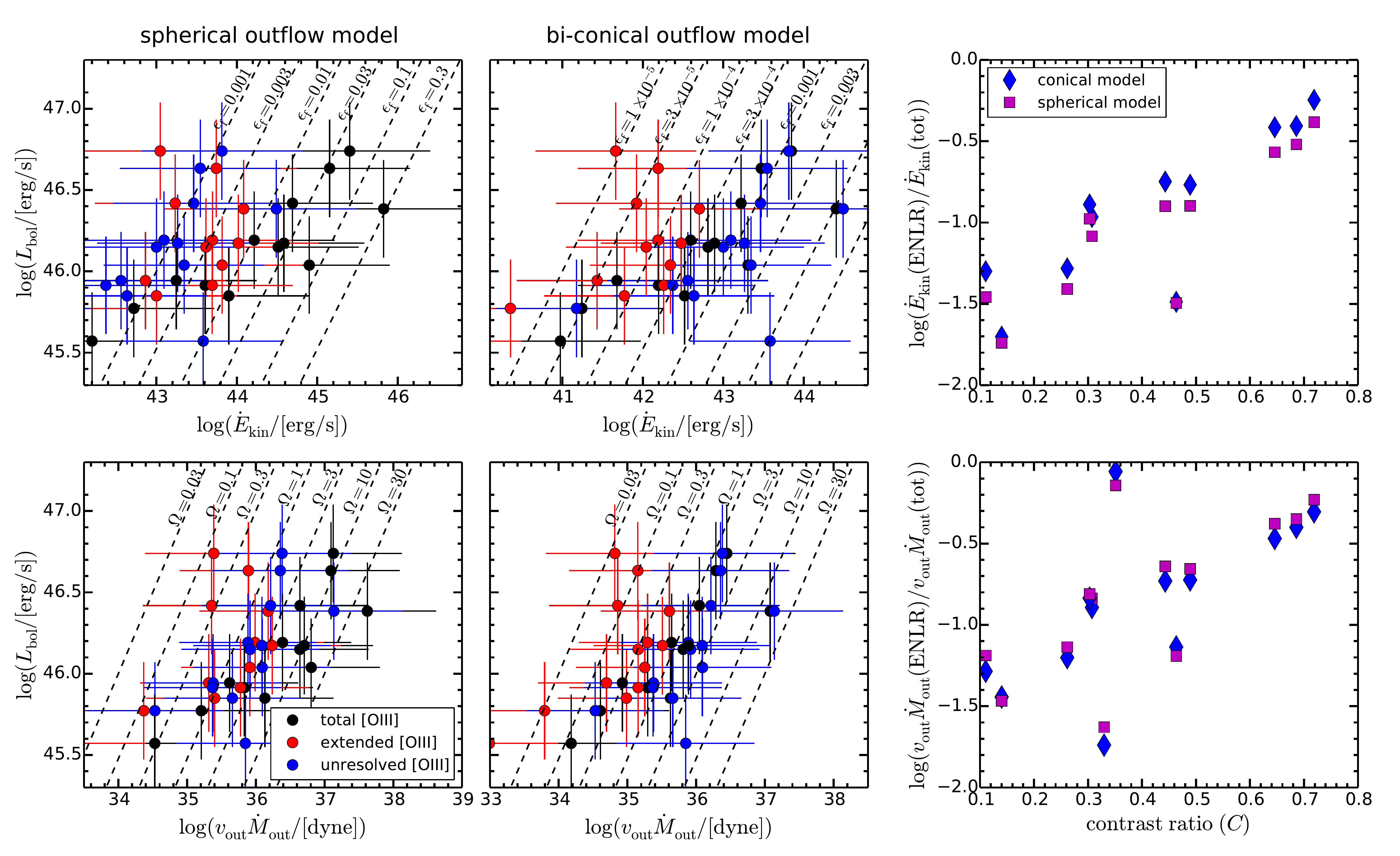}
\caption{Bolometric AGN luminosity ($L_\mathrm{bol}$) versus kinetic power ($\dot E_\mathrm{kin}$) and  momentum outflow rate 
($v_\mathrm{out}\dot M_\mathrm{out}$) for the spherical and conical outflow model at kpc-scale distances ($D=5\,\mathrm{kpc}$). The black and red data points correspond to the measurements 
from the total and extended [\OIII] line profile maps, respectively. In addition we also plot the corresponding values from the spatially unresolved [\OIII] line from the NLR as blue data 
points. This is based on the conical formula but also applies to the spherical case as described in the main text. The dashed lines correspond to various AGN feedback efficiencies ($\epsilon_f$) and 
outflow covering factors ($\Omega$) as labelled in the corresponding panels. The error bars on  $\dot E_\mathrm{kin}$ correspond to the range in electron densities $10<n_\mathrm{e}<1000$. The 
uncertainty on $L_\mathrm{bol}$ is assumed to be 0.3\,dex taking into account the photometric calibration of the data and the error on the bolometric correction factor. In the right panels, we show 
the change of the kinetic power and the momentum  outflow rate from the total and extended [\OIII] measurements as a function of contrast ratio ($C$).}
\label{fig:AGN_efficiency}
\end{figure*}

The asymmetric light distribution of [\OIII] emission together with the smaller line width on large scales questions the primary outflow mechanisms for these QSO. While 
\citet{Liu:2013b} proposed a wide-angle high-velocity outflow driven by the QSO radiation for these objects, \citet{Mullaney:2013} and \citet{Villar-Martin:2014} argued that the power of the radio 
jets is more strongly correlated with the line width than the AGN luminosity for a large sample of AGN from the SDSS (see, however, \citet{Woo:2016} for a different interpretation). Given the enhanced 
asymmetry in the [\OIII] light distribution it is unlikely that the proposed wide-angle radiation-driven outflow scenario is still valid for these QSOs. The ENLR morphologies that we recover rather 
support a conical geometry which is consistent with a preferred outflow axis as required by a radio-jet scenario. Given the redshift of the sample, the upper limits on the radio fluxes imply radio 
luminosities of $2\times10^{24}\,\mathrm{W\,Hz}^{-1}$. Although the QSOs are considered radio-quiet, these radio luminosities are consistent with those of 
low-redshift Seyferts and QSOs where broad emission lines on 100--1000\,pc scales could be directly associated with the hot spots of radio jets 
\citep[e.g.][]{Fu:2009,Mueller-Sanchez:2011,Husemann:2013a,Harrison:2015}. This matches with our findings that the very broad lines must be emitted on scales $<$1\,kpc that cannot be resolved with 
the seeing-limited optical observations at $z\sim0.6$. However, an alternative interpretation of the radio emission is that is generated by the shock front of an AGN-driven outflow itself 
\citep{Faucher-Giguere:2012,Zakamska:2014}.

Whether the impact of an unresolved NLR is similarly strong on the morphology of the ENLR and the outflow kinematics for the obscured QSOs is unclear. The similarity of the uncorrected 
maps inferred of the obscured and unobscured QSOs as discussed in \citet{Liu:2014} suggests that the beam smeared emission of an unresolved NLR affects both types similarly. However, 
a reliable deblending and estimation of the contrast between resolved and unresolved emission is difficult for the obscured QSOs without a simultaneous characterization of the 
PSF. Thus, we are not able to directly verify that effect for obscured QSOs directly with the existing observations, but \citet{Villar-Martin:2016} and \citet{Karouzos:2016} have recently 
reported also very compact outflow sizes of $<1-2$\,kpc after correction for a sample of luminous obscured QSOs at $z<0.6$. While a consistent picture of the systematic effects is appearing, we 
cannot entirely exclude that the ionized gas properties on kpc scales are intrinsically different in obscured and unobscured AGN given the different selection 
criteria used for the obscured QSO samples. 

\subsection{Implications for the AGN feedback efficiency}
A major goal of this study is to explore the impact of the beam smearing on the derived outflow energetics and mass outflow rates on large scales. We have shown that the kinetic power 
on kpc scales can drop up to two orders of magnitudes and the outflow rate up to one order of magnitude (see Table~\ref{tbl:energetics}), since the outflow velocity and line flux on kpc 
scales can be severely overestimated. Furthermore, we find a large difference in the results inferred for two popular outflow models that have frequently been used to infer AGN-driven large scale 
outflow energetics from integral-field spectroscopy data. This has significant impact on the AGN feedback efficiency $\epsilon_\mathrm{f}$, i.e. $\dot E_\mathrm{kin} 
=\epsilon_\mathrm{f}L_\mathrm{bol}$, which we show in Fig.~\ref{fig:AGN_efficiency} (upper left panels) for the spherical and the conical outflow model.  Here, we use the continuum luminosity at 
5100\AA\ as a proxy for $L_\mathrm{bol}$ with a bolometric correction factor of $L_\mathrm{bol}\sim10L_\mathrm{5100}$, following \citet{Richards:2006}. Errors are dominated by systematics in both 
cases and we adopt an order of magnitude error on the kinetic power due to the unknown electron density ($10<n_\mathrm{e}/\mathrm{cm}^{-3}<1000$) and an uncertainty of 0.3\,dex on the bolometric 
luminosity.

We find that the spherical outflow model reaches values up to $\epsilon_\mathrm{f}\sim0.3$ in the most extreme cases, when beam smearing is ignored. This decreases to a maximum value of 
$\epsilon_\mathrm{f}\sim0.01$, with a range of $0.001<\epsilon_\mathrm{f}<0.01$ after taking into account beam smearing from the unresolved NLR. Since the kinetic energies are much lower in 
the conical outflow model, we also compute lower AGN feedback efficiencies in the range of $10^{-5}<\epsilon_\mathrm{f}<10^{-4}$ after beam smearing correction. Similar feedback efficiencies have 
been inferred for outflows in nearby lower luminosity AGN that can be properly resolved \citep[e.g.][]{Barbosa:2009,Liu:2015}. The difference between the spherical and conical outflow models is 
important because current cosmological simulations including radiative QSO feedback predict or assume a feedback efficiency of  $\epsilon_\mathrm{f}\sim0.005$--$0.05$ 
\citep[e.g.][]{Matteo:2005,Hopkins:2010}. On the contrary, dedicated radiation-hydrodynamical simulations reported much lower feedback efficiencies in the range of 
$\epsilon_\mathrm{f}\sim10^{-8}$--$10^{-4}$ \citep{Kurosawa:2009}, but were computed on much smaller scales of 1--10\,pc and therefore do not quite match our spatial resolution. Thus, the answer to 
the question if AGN feedback is consistent with theoretical predictions depends strongly on the assumed theoretical model and outflow model as well as whether beam 
smearing is taken into account during the analysis of the data. 

To further check the reliability of the prediction of both models, we compute the momentum outflow rate $\dot P = \dot M v_\mathrm{out}$ based on our measurements provided in
Table~\ref{tab:comparison} and Table~\ref{tbl:energetics}. In the simple picture of a radiation-pressure driven wind, an upper limit is set by $\dot P < \Omega L_\mathrm{bol}/c$ where $\Omega$ is 
the covering factor and $c$ is the speed of light. In Fig.~\ref{fig:AGN_efficiency} (lower left panels), we compare the estimated momentum outflow rate with the corresponding limit based on the AGN 
bolometric luminosity of the QSO. We find that $\Omega>1$ in the spherical outflow model, even if beam smearing is considered. A covering factor larger than one can only be explained if the wind is 
not momentum conservative, or it is not radiatively driven in the first place, or the AGN luminosity has dropped on time scales much shorter than the dynamical time scale of the wind. For the conical 
outflow model, the observations including beam smearing correction are below the limit, with covering factors between $0.01<\Omega< 1$ as required for a simple radiatively driven QSO wind 
\citep[e.g.][]{Zubovas:2012,Stern:2016}. Thus, there would not need to be  an additional momentum boost to explain previous observations with $\Omega\gg1$ \citep[e.g.][]{Faucher-Giguere:2012} as beam 
smearing effects alone are able to solve this issue from the observational side. Recent observations of circum-nuclear winds that suggest $\dot P \sim \Omega L_\mathrm{bol}/c$ 
\citep[e.g][]{Tombesi:2015, Feruglio:2015} support our notion of an outflow that remains within the limit of radiation pressure.  Overall it is clear from Fig.~\ref{fig:AGN_efficiency} (right panels) 
that the overestimation of kinetic power and momentum outflow rate due to the beam smearing effect is strongly increasing with lower contrast ratio $C$ independent of the assumed model on the order of 
magnitude level.

The effect of beam smearing on the unresolved region NLR is the opposite as the light is spread out to large angular distances and the line width is slightly reduce by the blending from 
the large-scale quiescent kinematics. Thus, the associated energetics would be artificially reduced if simple aperture photometry is applied. It is therefore possible that 
most of the kinetic power and momentum outflow rate is actually confined to the unresolved NLR on scales $<1$\,kpc. This would imply a significant discontinuity of the mass outflow rate as function 
of radius, which may be a natural consequence of the finite outflow velocity combined with the finite life time of the bright AGN phase, of invalid assumptions for the physical conditions in 
the unresolved NLR, or of potential miss-interpretation of the unresolved NLR kinematics. Nevertheless, the estimated outflow rates in Table~\ref{tbl:energetics} for the unresolved NLR are 
still an order of magnitude smaller than the estimates for the large-scale wide-angle spherical outflow model of \citet{Liu:2013b} when beam smearing is not taken into account.

\begin{figure*}
\includegraphics[width=\textwidth]{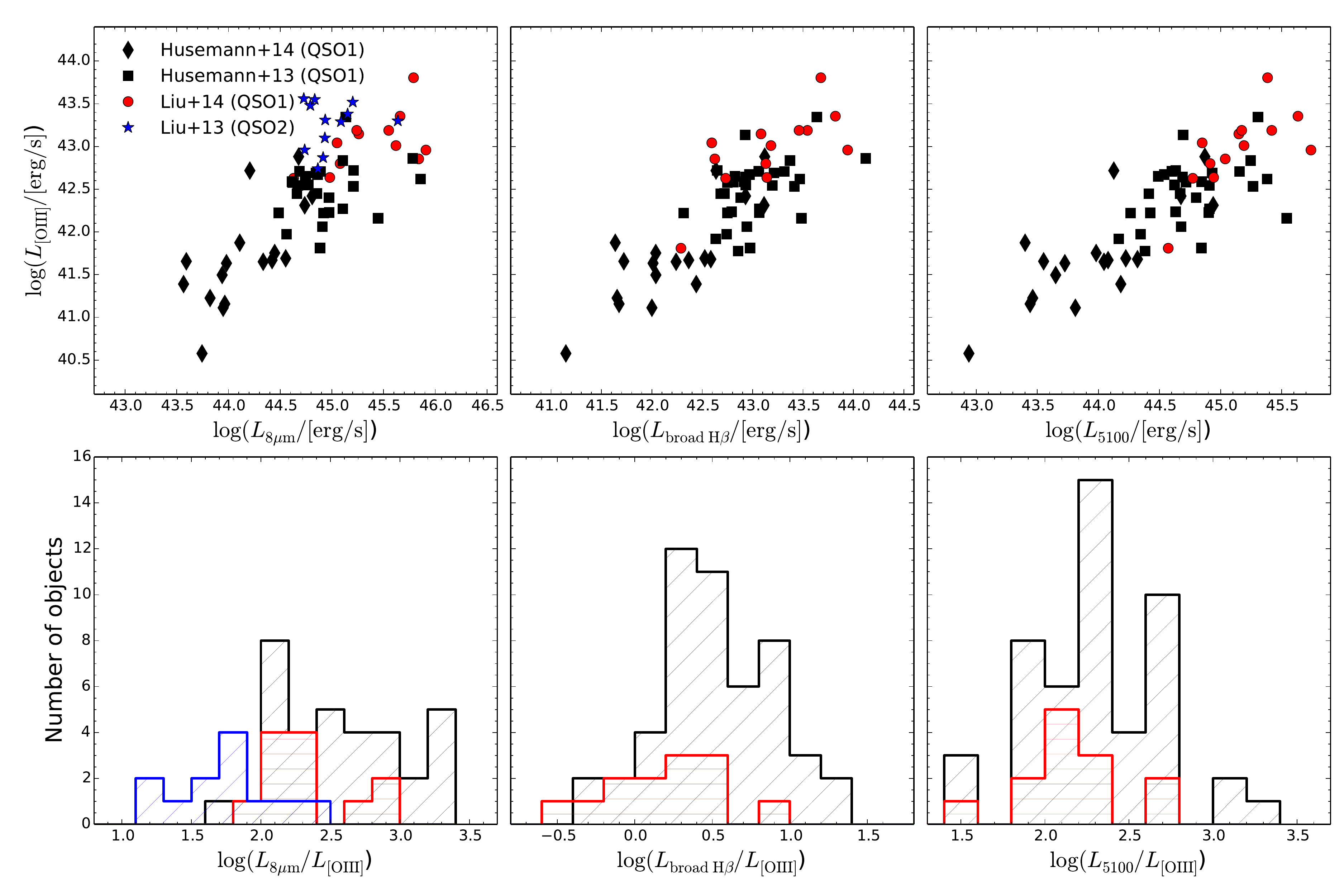}
\caption{Comparison of various AGN luminosity indicators are shown for the unobscured QSO samples of \citet{Husemann:2013a}, \citet{Husemann:2014} and \citet{Liu:2014} and for the obscured QSO sample 
of \citet{Liu:2013} in the upper panels. Here we consider the continuum luminosity at 8$\mu$m ($L_{8\mu\mathrm{m}}$), total [\OIII] luminosity ($L_\mathrm{[\OIII]}$), broad H$\beta$ luminosity  
($L_{\mathrm{broad\ H}\beta}$) and continuum luminosity at 5100\AA\ ($L_{5100}$) as independent AGN luminosity indicators. The corresponding distributions in the luminosity ratios are shown in the 
lower panels. For obscured QSOs only$L_\mathrm{[\OIII]}$ and  $L_{8\mu\mathrm{m}}$ can be measured because the broad lines from the BLR and the continuum from the accretion disc are obscured and 
unobservable. }
\label{fig:selection}
\end{figure*}

Besides a radiatively driven wind, it is also possible that a radio jet provides sufficient mechanical energy for powering a wind on galactic scales. Several studies have shown that high-velocity 
outflows seen as broad extended emission lines are co-spatial with kpc-scale jets even for radio-quiet AGN  \citep[e.g.][]{Fu:2009,Mueller-Sanchez:2011,Husemann:2013a}. Here, we compute an upper 
limit for the mechanical power of the putative radio jets from the upper limit in the 1.4\,GHz radio luminosity. The relation between jet ($P_\mathrm{jet}$) and mechanical power ($P_\mathrm{cav}$) by 
\citet{Cavagnolo:2010},
\begin{equation}
 \log\left(\frac{P_\mathrm{cav}}{10^{42}\,\mathrm{erg\,s}^{-1}}\right) = 0.75(\pm 0.14) \log\left(\frac{P_\mathrm{jet}}{10^{24}\,\mathrm{W\,Hz}^{-1}}\right)+1.91(\pm 0.18)
\end{equation}
yields an upper limit on the mechanical power of $P_\mathrm{cav}\lesssim8\times 10^{43}\,\mathrm{erg\,s}^{-1}$ based on the upper limit of $P_\mathrm{jet}\lesssim10^{24}\,\mathrm{W\,Hz}^{-1}$ valid 
for almost all QSOs in the sample. This limit is about an order of magnitude higher than the estimated kinetic power of the conical outflow model corrected for the beam smearing effect. Therefore, 
it is also possible, within the current observational limits, that the outflow in these QSOs is powered by low-luminosity radio jets inflating a cavity along the jet axis. If confirmed by deep 
high-resolution radio imaging, it would support the observation by \citet{Mullaney:2013} that the strength of the broad [\OIII] emission line component in stacked SDSS spectra of AGN is most strongly 
correlated with radio luminosity. 

In any case, the revised outflow energetics for this particular QSO sample suggest significantly weaker outflows than previously thought. With an ionized gas mass outflow rate of about $\dot 
M\sim10-100 M_\odot\,\mathrm{yr}^{-1}$ the QSOs would be able to expel the entire ionized gas content over a life time of 1--10\,Myr. Given that the total gas mass of the 
QSOs hosts and the mass-loading factor of the ionized gas outflows are unknown, it remains unclear whether QSO feedback in these cases is efficient enough to expel enough gas to significantly suppress 
star formation on short time-scales. Furthermore, it is still open whether the outflows are able to effect the entire host galaxies since the mass outflow rates in the unresolved compact NLR 
appear to be an order of magnitude higher than on kpc scales of the conical model.

\subsection{Are the results applicable also to obscured QSOs?}
\citet{Liu:2014} reported that the surface brightness distributions and kinematics are similar for the unobscured QSO and their matched obscured QSOs presented in \citet{Liu:2013,Liu:2013b}. 
Specifically, the ENLR morphology appeared to be round out to kpc scales with a high velocity dispersion in [\OIII] over a large region in both cases. This is already a surprising result as 
the AGN unification model \citep{Antonucci:1993} predicts that the inclination of the torus is different for unobscured and obscured QSOs and so we would expect the ionization cones oriented more  
perpendicular to our line-of-sight for obscured QSOs and pointing more towards us for unobscured QSOs. This projection effect should lead to different apparent morphologies of the ENLR from 
bi-conical for obscured and more round for unobscured QSOs.

The outflow energetics inferred by \citet{Liu:2013b} for the obscured QSOs is in agreement with our estimates for the matched unobscured QSO sample consistently assuming the spherical outflow 
geometry before corrected for the beam smearing effect of a compact NLR. Since we show that the beam smearing effect is prominent for the unobscured QSOs to recover the true morphology and 
kinematics of the ENLR on kpc scales, there are two possibilities to interpret the similarities reported by \citet{Liu:2014}: (i) The obscured QSOs are similarly affected by the beam smearing of a 
bright unresolved NLR outshining the ENLR, or (ii) the compact NLR close the nucleus is fainter or more strongly obscured by dust from the host galaxy in obscured QSOs compared to unobscured ones so 
that the beam smearing would be less problematic. 

To test whether the [\OIII] line in obscured QSOs is more strongly suppressed by dust obscuration we compare various AGN bolometric luminosities indicators for the different samples.
In Fig.~\ref{fig:selection}, we compare the [\OIII] luminosity ($L_\mathrm{[\OIII]}$) to the 8\,$\mu$m continuum luminosity ($L_{8\mu\mathrm{m}}$), the broad H$\beta$ luminosity 
($L_{\mathrm{broad\ H}\beta}$) and the continuum luminosity at 5100\,\AA\ ($L_\mathrm{5100}$) for the unobscured QSO sample discussed in this paper with the matched obscured QSO sample of 
\citet{Liu:2013}, and the large sample of 50 unobscured QSOs at $0.04<z<0.3$ \citep{Husemann:2013a,Husemann:2014}.
Here ,we use the [\OIII] luminosity as the primary AGN luminosity  reference given that the obscured and unobscured QSOs samples of \citet{Liu:2013,Liu:2014} were selected from the SDSS catalogue 
based on [\OIII] luminosity \citep{Reyes:2008}. The use of the [\OIII] luminosity as a bolometric luminosity indicator has been established over several order of magnitude 
\citep[e.g.][]{Zakamska:2003, Heckman:2004} which is the basis for the selection.

The unobscured QSOs studied in \citet{Husemann:2013a} and \citet{Husemann:2014} were selected from the Hamburg/ESO QSO survey \citep{Wisotzki:2000} and the Palomar Bright QSO survey 
\citep{Schmidt:1983} based on their continuum luminosity which is not obscured by the torus around the nucleus. For those QSOs, the continuum luminosity from the accretion disc and the broad 
H$\beta$ luminosity from the surrounding BLR are well calibrated AGN luminosity indicators that were shown to be directly linked through reverberation mapping 
\citep[e.g.][]{Kaspi:2000,Peterson:2004}. We find that all unobscured QSOs follow a clear correlation with the [\OIII] luminosity and that the unobscured QSOs have a similar distribution in the 
ratio with $L_\mathrm{5100}$ and $L_{\mathrm{broad\ H}\beta}$ as verified by a Kolmogorov-Smirnov test. 

Since the nucleus is obscured by the torus we cannot measure those quantities for the obscured QSOs. Thus, we use the infrared luminosity at $\sim$8\,$\mu$m as a measure for the AGN 
luminosity \citep[e.g.][]{RamosAlmeida:2007,Horst:2008,Gandhi:2009,Mateos:2015} as it is assumed to be the re-radiated dust emission from the AGN-heated torus. Indeed, we find the infrared 
luminosity to be closely correlated with the [\OIII] luminosity for the unobscured QSOs samples shown in Fig.~\ref{fig:selection} (left panel). However, we find a significantly different distribution 
of $L_{8\mu\mathrm{m}}/L_\mathrm{[\OIII]}$ for the obscured QSO sample of \citet{Liu:2013} compared to the unobscured QSO samples including the one matched in the [OIII] luminosity selection 
\citep{Liu:2014}. There appears to be an excess in [\OIII] compared to the AGN luminosity only for the obscured QSOs. This is likely a selection effect given that the sample is specifically selected 
to be the most luminous [\OIII] emitters at a given redshift picking up preferentially outliers.

Therefore, it is very unlikely that the NLR in the obscured QSOs is much weaker compared to the unobscured one. The opposite may even be true and the NLR is brighter relative to the ENLR in the 
obscured QSOs compared to the unobscured ones. Thus, a correct handling of the beam smearing effect seems to be at least as important for obscured  as for unobscured QSOs. The recent long-slit 
observations of obscured QSOs by \citet{Humphrey:2015} indeed confirm that a large fraction of the [\OIII] emission originates from a spatially unresolved component supporting our claim. We 
conclude that it is very likely that the effect of the beam smearing will need to be taken into account for the obscured QSOs and that the outflow energetics reported by \citet{Liu:2013b} are 
likely overestimated as well.

\section{Summary \& Conclusions}\label{sect:conclusion}
In this paper, we have presented an independent re-analysis of GMOS IFU spectroscopy of 12 luminous unobscured QSO at $0.4<z<0.7$ initially presented by \citet{Liu:2014}. We focus on 
the beam smearing effect associated with the spatially unresolved [\OIII] emission and its impact on the apparent large scale kinematics over several kpc and corresponding measures of AGN 
outflow energetics and feedback efficiencies. Our findings can be summarized as follows.
\begin{itemize}
 \item We find a large range in contrast ratios $0.2<C<1$ between the spatially resolved and total [\OIII] emission among the sample. The contrast ratio has a 
significant impact on the measured properties of the spatially extended [\OIII] emission on kpc scales. The radial surface brightness gradient has an intrinsically shallower power-law slope when the 
contribution from the unresolved component is removed.
\item While the estimated size of the ENLR is only overestimated at the lowest contrast ratios, we find an increasing asymmetry in the distribution of the truly extended [\OIII] emission
which is indicated by an offset in the flux-weighted centre with respect to the QSO position. More importantly, we notice that the [\OIII] line width ($W_{80}$) on kpc scales significantly decreases 
with decreasing contrast ratio after subtraction of the spatially unresolved [\OIII] component. 
\item We do not detect a clear break radius in the [\OIII]/H$\beta$ ratio for this unobscured QSO sample which is in stark contrast to the results for the obscured QSO sample presented by 
\citep{Liu:2013}. Together with the intrinsically asymmetric ENLR light distribution our observations provide evidence against the proposed spherical outflow model as the best interpretation of the 
observations.
\item The inferred AGN outflow energetics and feedback efficiencies on kpc scales dramatically decrease  with decreasing contrast ratio by up to $\sim$2 orders of magnitude in 
the in the most extreme case, when the unresolved emission of the QSO is taken into account. In addition, there is also a significant difference in the AGN feedback efficiency between the spherical 
and conical outflow model by 1-2 orders of magnitude. The (bi-)conical model seems to better match energetic constraints. 
\item The AGN outflow energetics and feedback efficiencies for the unresolved NLR  may carry most of the outflow power confined to a sub-kpc region 
with mass outflow rates of up to 100$M_\odot\,\mathrm{yr}^{-1}$. This is still an order of magnitude lower than in the spherical model without correction for beam smearing.
\item After correction for beam smearing the AGN feedback efficiencies reduce in all considered models to 0.01-0.1 per cent which are
lower efficiencies than assumed in many cosmological simulations to reproduce the galaxy population at high stellar masses. This could mean that either the kinetic coupling of the 
energy is small compared to thermal feedback or that current model assumptions have to be adjusted. In addition, we find that the momentum injection rate close or below the limit of 
$L_\mathrm{bol}/c$ for a radiatively-driven outflow in the conical case when beam smearing is taken into account. 
\end{itemize}

Overall our investigation implies that a proper handling of the beam smearing effect is crucial for interpreting and quantifying the energetics of AGN-driven outflows around luminous QSOs. This is 
particularly important for data providing low spatial resolution ($>$1\,kpc). The overestimation of AGN efficiencies is probably not just restricted to this specific QSO sample studied here, but may 
apply to other unobscured QSO studies at low and high redshift, in which beam smearing is not properly addressed yet. For example, it has not yet been discussed which role beam smearing plays in 
high-redshift QSOs showing signatures of kpc-scale broad [\OIII] lines \citep[e.g.][]{CanoDiaz:2012,Cresci:2015,Brusa:2015}. 
Of course, there are also individual examples of truly extended AGN outflows in objects where the light distribution of the broad [\OIII] line is clearly asymmetric on kpc 
scales \citep{Harrison:2015,Greene:2012}. While broad components in the [\OIII] line seem to be quite common for luminous AGN, which can be interpreted as AGN outflow, the beam smearing effect needs 
to be carefully evaluated to robustly quantify the large-scale kinematics on a case-by-case basis. 

Our results also imply that a significant part of the kinetic power and mass outflow rate may be confined to small scales ($<1$\,kpc) given the prominence of the unresolved NLR. It is 
difficult to verify these estimates without being able to verify if the assumption of constant-density clouds for the conical model actually holds or not. In any case, even with our conservative 
assumptions our estimates are more than an order of magnitude lower than in the case of the spherical outflow model if beam smearing is not taken into account.

Consequently, the AGN energetics can change drastically depending on the contrast ratio between the unresolved and resolved [\OIII] emission and its associated kinematics. Estimated AGN energetics 
depend strongly on (i) the measurements itself, given the contamination of extended emission by an unresolved component and (ii) assumptions made in the model, such as the outflow geometry and the 
electron density, if not measurable. For the QSO sample presented here, it means that the inferred kinematic power for the conical outflow model yields 1.5 to 3 orders of magnitude lower 
outflow energies compared to the spherical outflow model taking the unresolved emission contribution into account.

The beam smearing effect may therefore partially explain the discrepancy between the claim for ubiquitous kpc-scale outflows in obscured QSOs \citep{Liu:2013} and the contrary result for unobscured 
QSO by \citet{Husemann:2013a}.  However, it remains unclear whether the observation of unobscured and obscured QSOs are equally affected. A direct study of beam smearing effects for obscured QSOs is 
often infeasible but \citet{Villar-Martin:2016} and \citet{Karouzos:2016} have taken the beam smearing into account when analysing a sample of obscured QSOs and also report rather compact 
outflows size of 1--2\,kpc. Given that we find [\OIII] luminosity in obscured QSOs slightly enhanced compared to the AGN luminosity, the compact NLR region is not more obscured than in unobscured 
QSOs and makes the beam smearing effect as important as for unobscured QSOs. 

Therefore, we conclude that the question whether powerful kpc-scale outflows are ubiquitous in all AGN above a certain luminosity still remains open. Given that there is necessarily also a time 
evolution in the size of the outflow and its properties, there should be a population of luminous QSOs where the size of the outflow is still small and appears barely resolved. Such scenarios can 
only be studied when the beam smearing effect, ubiquitous for observational data, is taken into account.

\begin{acknowledgements}
We appreciate the valuable comments and suggestion of the referee that greatly improved the clarity and quality of the article. We thank Kevin Hainline for providing his measurements of the $8\mu$m 
luminosity for the QSO sample presented in \citet{Husemann:2013a} and Dominika Wylezalek for helping to retrieve the GMOS data 
from the Gemini archive.

JS acknowledges the European Research Council for the Advanced Grant Program 267399-Momentum.
VNB acknowledges assistance from a National Science Foundation (NSF) Research at Undergraduate Institutions (RUI) grant AST-1312296 and support for program number HST-AR-12625.11-A, provided by 
NASA through a grant from the Space Telescope Science Institute, which is operated by the Association of Universities for Research in Astronomy, Incorporated, under NASA contract NAS5-26555.
JHW acknowledges support by the National Research Foundation of Korea to the Center for Galaxy Evolution Research (2010-0027910).
Based on observations (Program ID: GN-2012B-Q-29) obtained at the Gemini Observatory acquired through the Gemini Science Archive, which is operated by the Association of Universities for 
Research in Astronomy, Inc., under a cooperative agreement with the NSF on behalf of the Gemini partnership: the National Science Foundation (United States), the National Research Council (Canada), 
CONICYT (Chile), the Australian Research Council (Australia), Minist\'{e}rio da Ci\^{e}ncia, Tecnologia e Inova\c{c}\~{a}o (Brazil) and Ministerio de Ciencia, Tecnolog\'{i}a e Innovaci\'{o}n 
Productiva (Argentina).

Funding for SDSS-III has been provided by the Alfred P. Sloan Foundation, the Participating Institutions, the National Science Foundation, and the U.S. Department of Energy Office of Science. The 
SDSS-III web site is http://www.sdss3.org/. SDSS-III is managed by the Astrophysical Research Consortium for the Participating Institutions of the SDSS-III Collaboration including the University of 
Arizona, the Brazilian Participation Group, 
Brookhaven National Laboratory, Carnegie Mellon University, University of Florida, the French Participation Group, the German Participation Group, Harvard University, the Instituto de Astrofisica de 
Canarias, the Michigan State/Notre Dame/JINA Participation Group, Johns Hopkins University, Lawrence Berkeley National Laboratory, Max Planck Institute for Astrophysics, Max Planck Institute for 
Extraterrestrial Physics, New Mexico State University, New York University, Ohio State University, Pennsylvania State University, University of Portsmouth, Princeton University, the Spanish 
Participation Group, University of Tokyo, University of Utah, Vanderbilt University, University of Virginia, University of Washington, and Yale University. 
\end{acknowledgements}

\bibliographystyle{aa}
\bibliography{references}

\end{document}

%% file: table1.tex
\begin{tabular}{lcccccccccccccc}\hline\hline
Identifier & $z$ & $\log L_\mathrm{OIII}$\tablefootmark{a} & $\log L_{\mathrm{H}\beta}$\tablefootmark{b} & $\log L_{5100}$\tablefootmark{c} & $ \log L_{8\mu\mathrm{m}}$\tablefootmark{d} & $f_\mathrm{1.4GHz}$ & $\log P_\mathrm{1.4GHz}$ & resolution\tablefootmark{e}\\
 & & [$\mathrm{erg\,s}^{-1}$] & [$\mathrm{erg\,s}^{-1}$] & [$\mathrm{erg\,s}^{-1}$] & [$\mathrm{erg\,s}^{-1}$] & [mJy] & [$\mathrm{W\,Hz}^{-1}$] & \\\hline
SDSS~J023342.57-074325.8 & $0.4538$ & $43.1$ & $42.6$ & $44.8$ & $45.0$ & $<1.1$ & $<23.8$ & $0.56''/3.2$\,kpc \\
SDSS~J030422.39+002231.8 & $0.6385$ & $42.8$ & $43.9$ & $45.7$ & $45.9$ & $<0.8$ & $<24.0$ & $0.55''/3.8$\,kpc \\
SDSS~J031154.51-070741.9 & $0.6330$ & $42.9$ & $42.6$ & $45.0$ & $45.8$ & $<1.1$ & $<24.1$ & $0.56''/3.8$\,kpc \\
SDSS~J041210.17-051109.1 & $0.5492$ & $43.5$ & $43.7$ & $45.4$ & $45.8$ & $3.2$ & $24.5$ & $0.48''/3.1$\,kpc \\
SDSS~J075352.98+315341.6 & $0.4938$ & $42.6$ & $42.7$ & $44.8$ & $44.6$ & $<1.0$ & $<23.9$ & $0.58''/3.5$\,kpc \\
SDSS~J080954.38+074355.1 & $0.6527$ & $43.2$ & $43.8$ & $45.6$ & $45.7$ & $<1.0$ & $<24.1$ & $0.67''/4.6$\,kpc \\
SDSS~J084702.55+294011.0 & $0.5662$ & $42.7$ & $43.1$ & $44.9$ & $45.0$ & $<1.0$ & $<24.0$ & $0.61''/4.0$\,kpc \\
SDSS~J090902.21+345926.5 & $0.5749$ & $43.1$ & $43.2$ & $45.2$ & $45.6$ & $<1.0$ & $<24.0$ & $0.68''/4.5$\,kpc \\
SDSS~J092423.42+064250.6 & $0.5884$ & $43.0$ & $43.5$ & $45.4$ & $45.5$ & $<1.1$ & $<24.1$ & $0.61''/4.0$\,kpc \\
SDSS~J093532.45+534836.5 & $0.6864$ & $43.2$ & $43.1$ & $45.1$ & $45.3$ & $<1.0$ & $<24.2$ & $0.77''/5.5$\,kpc \\
SDSS~J114417.78+104345.9 & $0.6785$ & $43.3$ & $43.5$ & $45.2$ & $45.2$ & $<1.0$ & $<24.2$ & $0.68''/4.8$\,kpc \\
SDSS~J221452.10+211505.1 & $0.4752$ & $42.8$ & $43.1$ & $44.9$ & $45.1$ & $<2.5$ & $<24.2$ & $0.46''/2.7$\,kpc \\
\hline\end{tabular}\tablefoot{\tablefoottext{a}{Total [OIII] line luminosity from \citet{Liu:2014}.}\tablefoottext{b}{Broad H$\beta$ line luminosity based on the QSO spectral modelling.}\tablefoottext{c}{QSO continuum luminosity at 5100\,\AA.}\tablefoottext{d}{Continuum luminosity at 8\,$\mu$m from \citet{Liu:2014}.}\tablefoottext{e}{Angular and physical resolution of the GMOS IFU data measured from the re-constructed broad H$\beta$ PSF.}}

%% file: table2.tex
\begin{tabular}{ccccccccccccccc}\hline\hline
Name & \multicolumn{2}{c}{$f_{\mathrm{[OIII]}}$\tablefootmark{a}} & $C$\tablefootmark{b} & \multicolumn{2}{c}{$d_\mathrm{QSO}$\tablefootmark{c}} & \multicolumn{2}{c}{$e$\tablefootmark{d}} & \multicolumn{2}{c}{$R_\mathrm{max}$\tablefootmark{e}} & \multicolumn{2}{c}{$\eta$\tablefootmark{f}} & \multicolumn{3}{c}{$W_{80}$\tablefootmark{g}}\\
 & \multicolumn{2}{c}{[$10^{-16}\,\mathrm{erg}\,\mathrm{s}^{-1}\,\mathrm{cm}^{-2}$]} & & \multicolumn{2}{c}{[kpc]} & \multicolumn{2}{c}{} & \multicolumn{2}{c}{[kpc]} & & & \multicolumn{3}{c}{[$\mathrm{km\,s}^{-1}$]}\\
 & NLR & ENLR &  & tot & ENLR & tot & ENLR & tot & ENLR & tot & ENLR & tot & ENLR & NLR\\\hline
SDSSJ0233-0743 & $154$ & $122$ & $0.44$ &  $0.199$ & $0.455$ & $0.18$ & $0.43$ & $17.1$ & $16.1$ & $2.1$ & $1.4$ & $471$ & $357$ & $568$\\
SDSSJ0304+0022 & $43$ & $21$ & $0.33$ &  $0.227$ & $1.402$ & $0.04$ & $0.20$ & $7.4$ & $7.4$ & $3.8$ & $3.1$ & $1307$ & $441$ & $1627$\\
SDSSJ0311-0707 & $53$ & $23$ & $0.31$ &  $0.331$ & $1.458$ & $0.08$ & $0.16$ & $10.0$ & $9.7$ & $4.0$ & $2.8$ & $1001$ & $774$ & $1072$\\
SDSSJ0412-0511 & $512$ & $83$ & $0.14$ &  $0.039$ & $0.553$ & $0.09$ & $0.11$ & $15.5$ & $18.3$ & $3.4$ & $2.0$ & $1204$ & $885$ & $1362$\\
SDSSJ0753+3153 & $43$ & $19$ & $0.30$ &  $0.090$ & $0.360$ & $0.05$ & $0.02$ & $7.1$ & $8.6$ & $3.9$ & $3.5$ & $259$ & $230$ & $265$\\
SDSSJ0809+0743 & $120$ & $42$ & $0.26$ &  $0.182$ & $2.366$ & $0.10$ & $0.24$ & $22.8$ & $22.8$ & $3.1$ & $2.0$ & $891$ & $693$ & $928$\\
SDSSJ0847+2940 & $19$ & $47$ & $0.72$ &  $0.750$ & $1.304$ & $0.03$ & $0.05$ & $12.0$ & $11.0$ & $2.8$ & $2.4$ & $362$ & $320$ & $910$\\
SDSSJ0909+3459 & $50$ & $110$ & $0.69$ &  $0.876$ & $1.777$ & $0.20$ & $0.18$ & $16.0$ & $14.2$ & $4.4$ & $4.1$ & $568$ & $512$ & $974$\\
SDSSJ0924+0642 & $85$ & $11$ & $0.11$ &  $0.100$ & $0.883$ & $0.06$ & $0.04$ & $9.5$ & $5.9$ & $3.5$ & $-0.5$ & $900$ & $765$ & $1064$\\
SDSSJ0935+5348 & $64$ & $61$ & $0.49$ &  $0.114$ & $0.364$ & $0.13$ & $0.26$ & $13.4$ & $14.0$ & $4.1$ & $3.3$ & $585$ & $490$ & $725$\\
SDSSJ1144+1043 & $63$ & $114$ & $0.65$ &  $0.384$ & $0.807$ & $0.23$ & $0.33$ & $20.2$ & $20.2$ & $3.6$ & $3.3$ & $676$ & $574$ & $900$\\
SDSSJ2214+2115 & $64$ & $35$ & $0.35$ &  $0.212$ & $1.125$ & $0.07$ & $0.02$ & $8.0$ & $9.3$ & $4.2$ & $3.9$ & $559$ & $835$ & $601$\\
\hline\end{tabular}\tablefoot{\tablefoottext{a}{Spatially integrated [OIII] line flux.}\tablefoottext{b}{Contrast ratio defined as $C=f_\mathrm{ENLR}/(f_\mathrm{NLR}+f_\mathrm{ENLR})$.}\tablefoottext{c}{Distance of the [OIII] flux-weighted centre with respect to the broad H$\beta$ flux-weighted centre defining the QSO position.}\tablefoottext{d}{Ellipticity of the [OIII] flux distribution.}\tablefoottext{e}{Maximum porjected size of the ENLR up to a local surface brightness of $\Sigma_\mathrm{[OIII]}>10^{-15}/(1+z)^{4}\,\mathrm{erg}\,\mathrm{s}^{-1}\,\mathrm{cm}^{-1}$.}\tablefoottext{f}{Power-law slope of the radial [OIII] surface brightness distribution between 1\arcsec--2.5\arcsec from the QSO.}\tablefoottext{g}{Median [OIII] line width as described in the text over a radius of $<$0.6\arcsec around the QSO.}}

%% file: table3.tex
\begin{tabular}{cccccccccccccccccccc}\hline\hline
      & $\log M_\mathrm{ion}$ & \multicolumn{2}{c}{$\log(E_\mathrm{kin}/[\mathrm{erg}])$}    &  \multicolumn{5}{c}{$\log(\dot E_\mathrm{kin}(D)/[\mathrm{erg}\,\mathrm{s}^{-1}])$\tablefootmark{a}} & \multicolumn{5}{c}{$\log(\dot M_\mathrm{out}(D)/[M_\odot\,\mathrm{yr}^{-1}])$\tablefootmark{b}} \\
Name      &  &\multicolumn{2}{c}{pixel-by-pixel} &  \multicolumn{2}{c}{spherical} &  \multicolumn{3}{c}{biconical} &  \multicolumn{2}{c}{spherical} &  \multicolumn{3}{c}{biconical}\\\cmidrule(l{5pt}r{5pt}){3-4}\cmidrule(l{5pt}r{5pt}){5-6}\cmidrule(l{5pt}r{5pt}){7-9}\cmidrule(l{5pt}r{5pt}){10-11}\cmidrule(l{5pt}r{5pt}){12-14}
 & tot & tot & ENLR & tot & ENLR & tot & ENLR & NLR & tot & ENLR &  tot &  ENLR  & NLR\\\hline
SDSSJ0233-0743 & $8.3$ & $55.3$ & $55.3$ & $43.9$ & $43.0$ & $42.5$ & $41.8$  & $42.6$ & $2.3$ & $1.7$ & $1.7$ & $1.1$ & $1.6$ \\
SDSSJ0304+0022 & $8.1$ & $56.8$ & $55.3$ & $45.4$ & $43.0$ & $43.8$ & $41.7$  & $43.8$ & $2.8$ & $1.6$ & $2.0$ & $0.9$ & $1.9$ \\
SDSSJ0311-0707 & $8.1$ & $55.5$ & $55.5$ & $44.9$ & $43.8$ & $43.3$ & $42.3$  & $43.3$ & $2.6$ & $1.9$ & $1.8$ & $1.1$ & $1.8$ \\
SDSSJ0412-0511 & $8.9$ & $55.9$ & $56.4$ & $45.8$ & $44.1$ & $44.4$ & $42.7$  & $44.5$ & $3.3$ & $2.2$ & $2.7$ & $1.4$ & $2.7$ \\
SDSSJ0753+3153 & $7.8$ & $54.4$ & $54.9$ & $42.7$ & $41.7$ & $41.2$ & $40.4$  & $41.2$ & $1.6$ & $0.9$ & $0.9$ & $0.2$ & $0.8$ \\
SDSSJ0809+0743 & $8.5$ & $56.2$ & $55.5$ & $45.2$ & $43.7$ & $43.5$ & $42.2$  & $43.5$ & $2.9$ & $1.9$ & $2.0$ & $1.0$ & $2.1$ \\
SDSSJ0847+2940 & $7.9$ & $54.2$ & $54.8$ & $43.2$ & $42.9$ & $41.7$ & $41.4$  & $42.6$ & $1.9$ & $1.7$ & $1.1$ & $0.9$ & $1.1$ \\
SDSSJ0909+3459 & $8.3$ & $54.7$ & $54.8$ & $44.2$ & $43.7$ & $42.6$ & $42.2$  & $43.1$ & $2.5$ & $2.2$ & $1.6$ & $1.3$ & $1.6$ \\
SDSSJ0924+0642 & $8.1$ & $55.6$ & $55.6$ & $44.7$ & $43.2$ & $43.2$ & $41.9$  & $43.5$ & $2.5$ & $1.4$ & $1.8$ & $0.7$ & $1.9$ \\
SDSSJ0935+5348 & $8.4$ & $55.6$ & $56.0$ & $44.5$ & $43.6$ & $42.8$ & $42.0$  & $43.0$ & $2.7$ & $2.1$ & $1.7$ & $1.2$ & $1.8$ \\
SDSSJ1144+1043 & $8.6$ & $55.9$ & $56.0$ & $44.6$ & $44.0$ & $42.9$ & $42.5$  & $43.3$ & $2.7$ & $2.4$ & $1.8$ & $1.5$ & $1.8$ \\
SDSSJ2214+2115 & $7.9$ & $54.9$ & $55.6$ & $43.6$ & $43.7$ & $42.2$ & $42.3$  & $42.4$ & $2.0$ & $1.8$ & $1.3$ & $1.0$ & $1.3$ \\
\hline\end{tabular}\tablefoot{\tablefoottext{a}{Kinetic power dreived from the total light and ENLR distribution for distance $D$=5\,kpc from the QSO for the spherical and conical model. For the conical model we also estimate the kinetic power for the unresolved NLR adopting a distance of $D$=1kpc.}\tablefoottext{b}{Mass outflow rate from the total light and ENLR distribution for distance $D$=5\,kpc from the QSO for the spherical and conical model. For the conical model we also estimate the kinetic power for the unresolved NLR adopting a distance of $D$=1kpc.}}